\documentclass[sigplan,nonacm]{acmart}

\usepackage{algorithm}
\usepackage{algpseudocode}
\usepackage{caption}
\usepackage{booktabs}
\usepackage{array}
\usepackage{tabularx}
\usepackage{amsmath}
\usepackage{graphicx}
\usepackage[]{hyperref}
\usepackage{xspace}
\usepackage{braket}
\usepackage{balance}
\usepackage{amsfonts}
\usepackage{subfig}
\usepackage{booktabs}
\usepackage{soul}
\usepackage{xcolor}

\newcommand{\ours}{Red-QAOA\xspace}
\newcommand{\nodered}{28\%\xspace}
\newcommand{\edgered}{37\%\xspace}
\newcommand{\mse}{2\%\xspace}

\ccsdesc[500]{Computer systems organization~Quantum computing}
\ccsdesc[300]{Mathematics of computing~Simulated annealing}
\ccsdesc[500]{Mathematics of computing~Combinatorial optimization}

\copyrightyear{2024}
\acmYear{2024}
\setcopyright{licensedusgovmixed}\acmConference[ASPLOS '24]{29th ACM International Conference on Architectural Support for Programming Languages and Operating Systems, Volume 2}{April 27-May 1, 2024}{La Jolla, CA, USA}
\acmBooktitle{29th ACM International Conference on Architectural Support for Programming Languages and Operating Systems, Volume 2 (ASPLOS '24), April 27-May 1, 2024, La Jolla, CA, USA}
\acmDOI{10.1145/3620665.3640363}
\acmISBN{979-8-4007-0385-0/24/04}

\title{Red-QAOA: Efficient Variational Optimization\\through Circuit Reduction}

 \author{Meng Wang}
 \orcid{0009-0008-1749-7929}
 \affiliation{%
   \institution{The University of British Columbia (UBC) and PNNL}
   \city{Vancouver}
   \country{Canada}}
 \email{mengwang@ece.ubc.ca}

 \author{Bo Fang}
 \orcid{0000-0001-9721-3982}
 \affiliation{%
   \institution{Pacific Northwest National Laboratory (PNNL)}
   \city{Richland}
   \country{USA}}
 \email{bo.fang@pnnl.gov}

 \author{Ang Li}
 \orcid{0000-0003-3734-9137}
 \affiliation{%
   \institution{Pacific Northwest National Laboratory (PNNL)}
   \city{Richland}
   \country{USA}}
 \email{ang.li@pnnl.gov}

 \author{Prashant J. Nair}
 \orcid{0000-0002-1732-4314}
 \affiliation{%
   \institution{The University of British Columbia (UBC)}
   \city{Vancouver}
   \country{Canada}}
 \email{prashantnair@ece.ubc.ca}

\renewcommand{\shortauthors}{Wang et al.}

\begin{document}
\begin{abstract}

The Quantum Approximate Optimization Algorithm (QAOA) addresses combinatorial optimization challenges by converting inputs to graphs. However, the optimal parameter searching process of QAOA is greatly affected by noise. Larger problems yield bigger graphs, requiring more qubits and making their outcomes highly noise-sensitive. This paper introduces Red-QAOA, leveraging energy landscape concentration via a simulated annealing-based graph reduction.

Red-QAOA creates a smaller (\emph{distilled}) graph with nearly identical parameters to the original graph. The distilled graph produces a smaller quantum circuit and thus reduces noise impact. At the end of the optimization, Red-QAOA employs the parameters from the distilled graph on the original graph and continues the parameter search on the original graph. Red-QAOA outperforms state-of-the-art Graph Neural Network (GNN)-based pooling techniques on 3200 real-world problems. Red-QAOA reduced node and edge counts by 28\% and 37\%, respectively, with a mean square error of only 2\%. 

\end{abstract}
\maketitle 
\pagestyle{plain} 

\begingroup
    \renewcommand\thefootnote{}%
    \footnotetext{© 2024 Copyright held by the owner/author(s). Publication rights licensed to ACM. This is the author's version of the work. It is posted here for your personal use. Not for redistribution. The definitive Version of Record was published in Proceedings of the 29th ACM International Conference on Architectural Support for Programming Languages and Operating Systems, Volume 2, http://dx.doi.org/10.1145/3620665.3640363.}%
    \addtocounter{footnote}{-1}%
\endgroup

\section{Introduction}
\label{sec:intro}
Quantum computing, particularly with Noisy Intermediate-Scale Quantum (NISQ) computers, offers a powerful solution for tackling complex algorithms~\cite{supremacy, Bruzewicz2019, Cao2019, Farhi2001, Feynman1982, ibm_roadmap, grover_fast_1996, shor_polynomial-time_1999, dasmultiprogramming}. The Quantum Approximate Optimization Algorithm (QAOA), a widely recognized Variational Quantum Algorithm (VQA), addresses intricate optimization problems in graph theory, supply chain optimization, and machine learning~\cite{farhi_quantum_2014, fletcher_practical_2000, lotshaw_empirical_2021, wang_quantum_2018, ward_qaoa_2018, zhou_quantum_2020,cerezo_variational_2021, endo_variational_2020, huang_near-term_2022}. To this end, QAOA treats inputs as graphs and maps nodes of the graph into qubits. However, modern NISQ machines, being inherently noisy, struggle to provide meaningful outcomes for large graphs as they use larger amounts of qubits~\cite{mengquantumsim,wang2022tqsim}. This leads to deep circuits that require a substantial number of qubits. Additionally, the execution of larger graphs tends to prolong processing times, reducing the overall throughput of the NISQ machine. This paper seeks to enhance the execution of larger graphs while ensuring meaningful outcomes.

Our paper examines QAOA applied to the MaxCut problem in graphs. These problems are NP-hard and are practically important. QAOA implementations harness the synergy between classical and NISQ computers. The classical computer furnishes quantum circuit parameters, while the NISQ computer maps the quantum circuit's graph nodes to qubits, executes it with the provided parameters, and generates optimization outcomes. The classical computer then updates circuit parameters based on these outcomes. This iterative feedback loop continues until the solution converges on optimal parameters~\cite{mengrestart}. The optimal parameters are then used to compute the Max-Cut of the graph. Ideally, regardless of the size of the input graph, the QAOA implementation should seamlessly operate on NISQ computers. However, practical NISQ machines encounter two critical concerns.

\begin{figure}[t]
    \centering
    \includegraphics[width=\columnwidth]{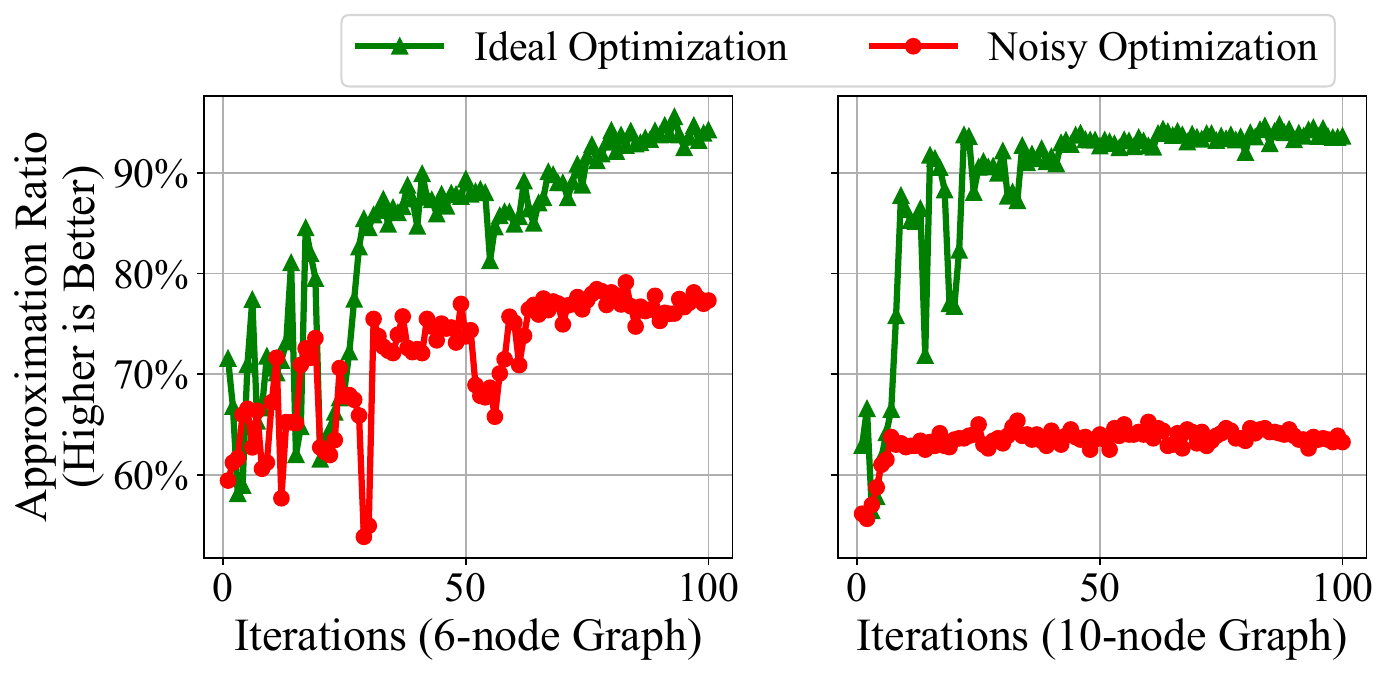}
    \caption{Comparing QAOA MaxCut approximation ratios (indicating outcome proximity to ground truth) between two graphs (6-node and 10-node) with noisy and ideal optimization. We perform 100 iterations for this comparison. The comparison illustrates (a) divergence from the ideal scenario as iterations increase (left) and (b) stagnating approximation ratios when scaling from 6-node to 10-node graphs (right).}
    \label{fig:motivation_error}
    \vspace{-0.1in}
\end{figure}

\noindent \textbf{1. Noise-Induced Degradation:} The approximation ratio, determining the closeness of QAOA outcomes to the ground truth, experiences significant degradation due to noise in the system. This is particularly high for larger problem instances. Figure~\ref{fig:motivation_error} illustrates the QAOA convergence rate for 6-node and 10-node graphs. When an ideal optimizer is employed, the approximation ratio steadily increases for both graphs, surpassing 90\%, indicating excellent performance. However, with noisy optimization (NISQ), the QAOA optimization's approximation ratio is markedly affected as the number of iterations rises. Thus, the cumulative nature of noise influences the \emph{usability} of NISQ computers~\cite{tannuecc}. Additionally, as the graph size scales from 6 nodes to 10 nodes, the required number of qubits also increases. As the impact of noise becomes severe at increased qubits, the approximation ratio for the 10-node graph remains stagnant at around 60\%. In contrast, the smaller 6-node graph maintains a relatively higher ratio of approximately 80\%. This highlights the heightened impact of noise for larger QAOA problem instances. 

\begin{figure}[b]
    \centering
    \includegraphics[width=\columnwidth]{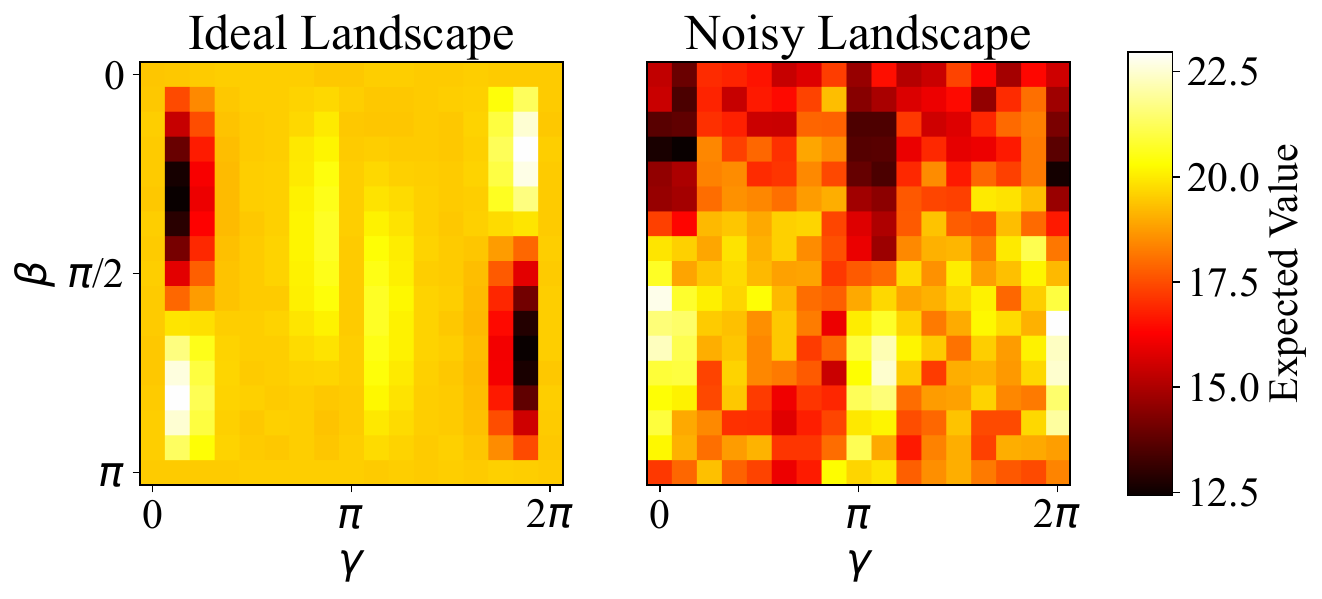}
    \caption{A comparison of the ideal QAOA energy landscape (left) and the noisy landscape executed on the 27-qubit \emph{ibmq\_kolkata} system (right) for a 13-node graph. We can observe that noise-induced distortions result in significant differences between the two energy landscapes.}
    \label{fig:motivation_landscape}
\end{figure}

\noindent \textbf{2. Distorted Solution Space:} Noise distorts the underlying energy landscape~\cite{tannunisq,tannumeasurement,tannudissimilar}. It actively misguides the optimization process and results in suboptimal outcomes. Figure~\ref{fig:motivation_landscape} compares the \emph{ideal QAOA energy} landscape (left) with the noisy landscape executed on the 27-qubit \emph{ibmq\_kolkata} system (right) for a 13-node graph. Figure~\ref{fig:motivation_landscape} highlights the substantial differences from noise-induced distortions. 

\noindent \textbf{State-of-art approaches:} One may substitute a large QAOA problem with an equivalent smaller one sharing the same optimal parameters. This leverages QAOA's unique property of similar instances having comparable energy landscapes~\cite{brandao_for_2018, farhi_quantum_2014, shaydulin_qaoakit_2021}. Prior work has explored the practicality of transferring optimal parameters between graphs~\cite{brandao_for_2018, galda_transferability_2021, shaydulin_parameter_2023}. However, existing frameworks encounter scalability hurdles due to stringent preconditions, limiting transferability to small subgraphs. For larger graphs, parameters must be transferable across multiple subgraphs within the larger graph (donor graphs), which acceptor graphs can then utilize. These approaches have two limitations: 1) Identifying and transferring parameters for numerous subgraph combinations is computationally hard, and 2) strict mutual transfer conditions for large graphs restrict their practicality.

\noindent \textbf{Our Proposal:} We present \ours, a novel simulated annealing-based approach designed to overcome these limitations. \ours generates a reduced (distilled) graph for optimal parameter search. Once the parameters are determined using this \emph{single distilled graph}, they are transferred back to the original graph. Subsequently, the optimization process proceeds using the original graph to converge on its precise optimal parameters. \ours ensures a less error-prone and more efficient optimization process. This is because using a distilled graph enables executing a smaller quantum circuit for a significant portion of the optimization.

\vspace{0.05in}
\noindent This paper makes four key contributions:
\begin{itemize}
\renewcommand{\labelitemi}{\scriptsize$\blacksquare$}
\item We propose \ours, a novel approach to search for optimal parameters with a reduced circuit for QAOA. \ours addresses noise challenges and outperforms state-of-the-art graph reduction methods.
\item We comprehensively examine the theoretical definition of similar instances in QAOA. Subsequently, we integrate a dynamic simulated annealing-based graph reduction algorithm within \ours to enable a more generalized approach.
\item We showcase the effectiveness of \ours by achieving a reduction of \nodered nodes and \edgered edges, along with a low Mean Squared Error (MSE) of \mse between ideal energy landscapes. These results are demonstrated across both real-world datasets and random graphs.
\item We assess \ours on real quantum devices, demonstrating substantial improvements over noisy baselines. Moreover, \ours seamlessly integrates with existing optimization methods, enhancing the performance of quantum optimization algorithms.
\end{itemize}

\section{Background}

\subsection{Basics of Quantum Computing}

Quantum computing is a computational paradigm that leverages the principles of quantum mechanics to perform calculations. The fundamental building blocks of quantum computers are qubits, quantum gates, and quantum circuits.

\subsubsection{Qubits}

Unlike classical bits, which can take the value of either 0 or 1, a qubit can exist as a linear combination of its basis states, $|0\rangle$ and $|1\rangle$, as shown:

\begin{equation}
|\psi\rangle = \alpha |0\rangle + \beta |1\rangle,
\end{equation}

where $\alpha$ and $\beta$ are complex coefficients, subject to the normalization condition $|\alpha|^2 + |\beta|^2 = 1$.

\subsubsection{Quantum Operations: Gates and Circuits}

Quantum gates are unitary transformations used to manipulate the states of qubits. Common examples include the Pauli-X, Y, Z, Hadamard, and CNOT gates. For instance, the Pauli-X gate is represented by the following matrix:

\begin{equation}
X = \begin{pmatrix}
0 & 1 \\
1 & 0
\end{pmatrix}.
\end{equation}

When the Pauli-X gate acts on a qubit $|\psi\rangle$, it flips the qubit's state such that $X|0\rangle = |1\rangle$ and $X|1\rangle = |0\rangle$.

A quantum circuit is a sequence of gates applied to a set of qubits~\cite{dasimitation,dasadapt}. The circuit represents a computation that transforms an input state into an output state.

\subsection{Variational Quantum Algorithms}

Variational Quantum Algorithms (VQAs) are a class of quantum algorithms that aim to solve optimization problems using quantum computers~\cite{cerezo_variational_2021,endo_variational_2020,huang_near-term_2022,dangwalvarsaw}. They harness quantum parallelism and interference to search for optimal solutions efficiently. VQAs use a parameterized quantum circuit to find the optimal solution. This is done by minimizing a cost function using classical optimization techniques.

\subsubsection{Quantum Approximate Optimization Algorithm}
Quantum Approximate Optimization Algorithm (QAOA)~\cite{farhi_quantum_2014} is a popular VQA. QAOA is designed to solve combinatorial optimization problems~\cite{tannuhammer}. These problems involve finding the best arrangement or order of objects or variables given certain constraints or criteria. QAOA aims to maximize the expected value of the cost Hamiltonian $H_c$ and the mixer Hamiltonian $H_m$ with respect to a trial state $\ket{\psi(\boldsymbol{\gamma}, \boldsymbol{\beta})}$ obtained from a parameterized quantum circuit. The trial state is produced by applying \textit{p} alternating layers of unitary operators (called QAOA layers) as shown in Equation~\ref{eq:trail_state}:
\begin{equation}
\label{eq:trail_state}
U(\boldsymbol{\gamma}, \boldsymbol{\beta}) = e^{-i\beta_p H_m} e^{-i\gamma_p H_c} \cdots e^{-i\beta_1 H_m} e^{-i\gamma_1 H_c} \ket{s}
\end{equation}

where $\ket{s}$ is the uniform superposition over
computational basis states shown in Equation~\ref{eq:init_state},
\begin{equation}
\label{eq:init_state}
\ket{s} = \frac{1}{\sqrt{2^n}} \sum_{z} \ket{z}
\end{equation}

where $n$ is the number of qubits. For the Maxcut problem, the cost Hamiltonian $H_c$ and the mixer Hamiltonian $H_m$ can be defined as follows:
\begin{equation}
H_c = \sum_{\langle i, j \rangle \in E} \frac{1}{2} (I - \sigma_i^z \sigma_j^z),
\end{equation}
\begin{equation}
H_m = \sum_{i=1}^{n} \sigma_i^x,
\end{equation}

where $\sigma_i^z$ and $\sigma_i^x$ represent the Pauli-Z and Pauli-X operators acting on the $i$-th qubit, respectively, and $E$ signifies the set of edges in the input graph. The Maxcut solutions are encoded in the eigenstates of the problem Hamiltonian, where each qubit denotes a vertex of the graph, with its state indicating the partition (0 or 1) to which the vertex belongs.

\subsubsection{Graph Reduction for QAOA}

QAOA operates on graphs, so reducing the problem size entails shrinking the underlying graph representation. This reduction effectively transforms the circuit reduction problem into a graph reduction problem. Graph reduction is a well-explored field with techniques that simplify graphs while preserving their structural or functional properties~\cite{deo_graph_2016}.

Graph neural network (GNN)-based graph pooling methods offer promising techniques for graph reduction while maintaining graph structures. Attention-based Spectral Aggregative (ASA) pooling~\cite{ranjan_asap_2020} and Self-Attention Graph (SAG) pooling~\cite{knyazev_understanding_2019, lee_self-attention_2019} utilize attention mechanisms to learn reduced representations of graphs. ASA pooling aggregates node features across scales using attention coefficients and multiscale Laplacian eigenvectors. In contrast, SAG pooling learns soft hierarchical node clustering by computing importance scores for each node. Another approach, Top-K pooling~\cite{cangea_towards_2018, gao_graph_2019, knyazev_understanding_2019}, is a pooling method that selects the top-K nodes based on learned importance scores, constructing a smaller graph retaining the most relevant information.
\section{Motivation: Equivalent Graph Instances}
VQA executions involve two primary steps to solve the underlying optimization problem: parameter optimization and solution-finding. The first step aims to determine the optimal parameters of a parameterized quantum circuit to minimize a cost function. On the other hand, the second step, called solution-finding, seeks to find the optimal solution using the parameters obtained from the previous step.

Prior works have focused on finding equivalent graph instances for the solution-finding step. However, we argue that one can more efficiently optimize QAOA by finding equivalent instances for parameter optimization.

\subsection{Parameter Optimization}
The parameter optimization step is significantly more prone to errors and requires more computational resources. It entails employing classical optimization algorithms, which involve iteratively evaluating circuits with varying parameter values on a NISQ machine. Errors within qubits may accumulate during each circuit evaluation due to noise and imperfections inherent in the quantum (NISQ) hardware. These accumulated errors can potentially impede the accuracy of the parameter optimization process.

Moreover, the optimization process might entail navigating through a vast parameter space, increasing the likelihood of encountering local minima—suboptimal solutions that are not the global minimum. This situation arises when the optimization algorithm becomes trapped within a specific region of the parameter space, neglecting exploration of other regions that could potentially harbour superior solutions. Furthermore, the combination of complex quantum circuits with the iterative classical optimization process makes the parameter optimization step more prone to errors.

\subsection{Challenge: Finding Similar Graphs}
The definition of equivalent instances differs between the parameter optimization and solution-finding steps in VQA due to their distinct objectives. In the solution-finding step, instances are considered equivalent if their outcomes are \emph{exactly} the same~\cite{Lotshaw_2021}. This poses challenges, especially for large graphs, as it requires converging on a few reduced graphs that yield identical outcomes. 

In contrast, instances are considered equivalent in the parameter optimization step if they share \emph{similar} (or nearly-identical) optimal parameters that minimize the cost function~\cite{farhi_quantum_2014, brandao_for_2018, galda_transferability_2021}. This broader criterion allows for operating on significantly more reduced graphs in this step. Thus, the equivalence requirements in parameter optimization are more flexible than the solution-finding step. This relaxed criterion enables the selection of instances that may not be exactly identical but possess similar properties, enabling more efficient optimization of larger problem instances.

\subsection{Observation: Common Energy Landscape}
\label{sec:transfer_theory}
When addressing the Maxcut problem for a specific graph using QAOA, the operators in Equation~\ref{eq:trail_state} can be partitioned into sub-terms. Each sub-term corresponds to an edge within the graph. By commuting the terms, the resulting subterm operators involve only qubits $j$ and $k$ and any other qubits with a graph distance from $j$ or $k$ no greater than $p$ layers~\footnote{For a more in-depth discussion, we refer the reader to the seminal QAOA paper by Farhi et al.~\cite{farhi_quantum_2014}.}.

The expected value can be expressed as a sum of sub-expected values, each calculated on a subgraph. The subterm operators are also identical when comparing two graphs with identical subgraphs. Consequently, the overall QAOA for both graphs is indistinguishable, allowing for the direct transfer of optimal parameters from one graph to the other. Not only are the optimal parameters transferable in such cases, but the entire energy landscapes of the two graphs also coincide. In the context of QAOA, an energy landscape refers to the distribution of energy values (or objective function values) over possible solutions or configurations.

\begin{figure}[t]
    \centering
    \includegraphics[width=0.9\columnwidth]{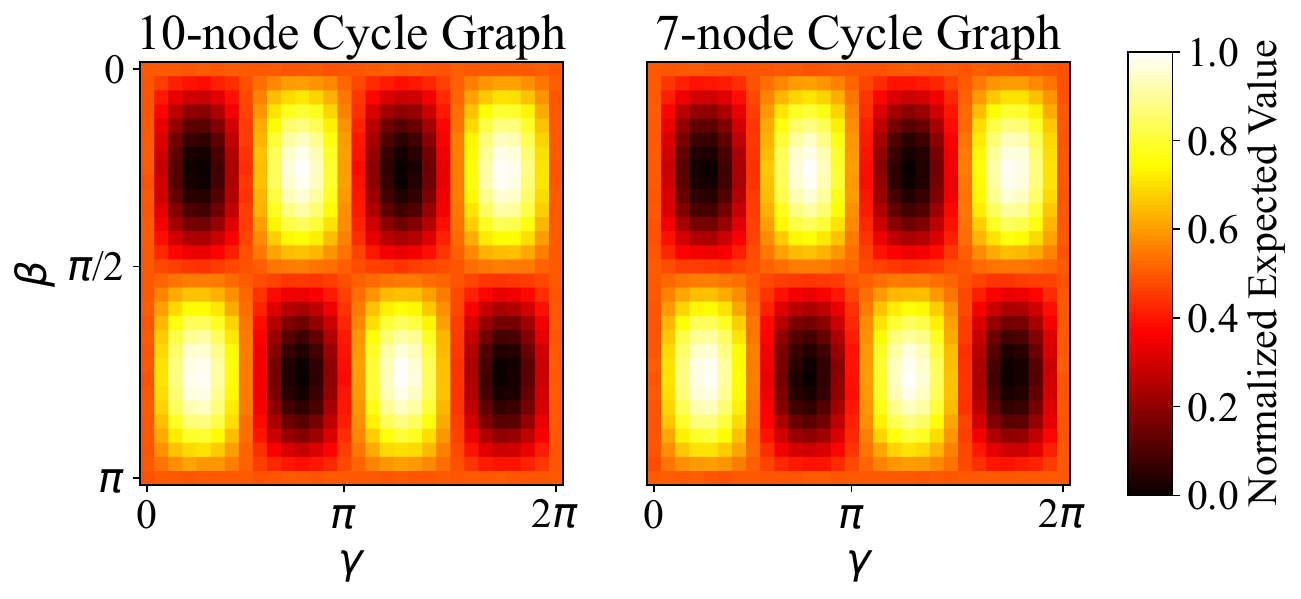}
    \caption{QAOA energy landscapes for 7-node and 10-node cycle graphs. $\beta$ and $\gamma$ are in the ranges of [0, $\pi$] and [0, 2$\pi$], respectively.  Since cycle graphs share the same sub-graphs, they exhibit nearly identical energy landscapes.}
    \label{fig:identical_landcape}
\end{figure}

Figure~\ref{fig:identical_landcape} presents two energy landscapes acquired using 7- and 10-node cycle graphs. A cycle graph comprises a single closed loop of nodes and edges, with each node connecting to two other nodes. As a result, irrespective of the number of nodes, they always share the same subgraphs, leading to almost identical energy landscapes in the two cases.

\begin{figure*}
    \centering
    \includegraphics[width=0.85\linewidth]{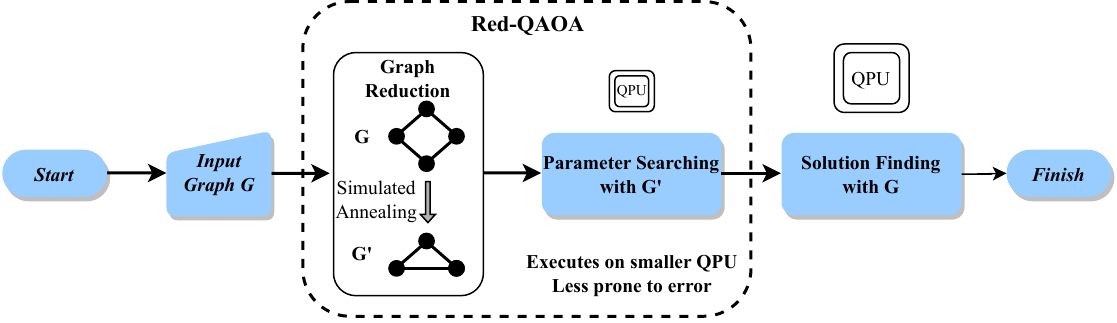}
    \caption{High-level overview of the design of \ours. The dashed block indicates the main component of \ours, which includes a graph reduction step and optimal parameter searching with the reduced graph ($G'$). The reduced graph can be executed on a smaller quantum processing unit (QPU) and is less prone to errors. For the final step, we find the solution with the optimal parameter and the original graph ($G$). This step is more prone to error as we execute this on a larger QPU. However, since we only need to execute this only for final optimal and more accurate parameters, we can apply error mitigation to improve the overall execution accuracy.}
    \label{fig:design_overview}
\end{figure*}

\subsection{Idea: Leverage Energy Landscape Similarity}

When searching for optimal parameters for QAOA, we leverage the observation that we can substitute one energy landscape with another if the two landscapes are nearly identical. This holds even if the landscape is from a reduced graph. Thus, our key goal is to find the reduced graph efficiently.

We assess the similarity of the two energy landscapes' shapes by calculating the mean square error (MSE) between their normalized versions. Normalizing the energy landscapes ensures their energy ranges are comparable. Equation~\ref{eq:mse} presents the MSE calculation, discussed in more detail in Section~\ref{sec:merit}. A small value of MSE implies that the two landscapes have a similar shape, while a large value of MSE suggests a significant difference in shape. In Figure~\ref{fig:identical_landcape}, the MSE between the normalized energy landscapes is $1.6 \times 10^{-5}$, indicating that they are nearly identical.

\section{Our Proposal: \ours}

We propose \ours, a framework that addresses the challenges of noise and execution overhead in QAOA. \ours seeks reduced graphs that maintain the energy landscape of the original graph for efficient quantum computation.

Figure~\ref{fig:design_overview} presents a high-level overview of the \ours design. The main components, depicted in the dashed block, include the \textbf{graph reduction step} and \textbf{optimal parameter searching step}—using the reduced graph ($G'$). \ours minimizes errors and improves accuracy by executing the reduced graph on a smaller quantum processing unit (QPU). After determining the \emph{final optimal parameters}, \ours executes the original graph ($G$) on a larger QPU. This allows the application of error mitigation techniques on $G$, as $G$ executes only for the \emph{final optimal and accurate parameters}.

Our design consists of two main components: (1) First, we identify the key metric for selecting smaller equivalent instances, and (2) Second, we develop a graph reduction method based on this identified metric. 

\subsection{Theoretical Foundation of Red-QAOA}

The energy corresponding to QAOA parameter values can be expressed as a sum of local energies as shown in Equation~\ref{eq:local_energy_qaoa}:

\begin{equation}
E(\gamma,\beta) = \sum\limits_{<jk>} E_{<jk>}(\gamma,\beta)
\label{eq:local_energy_qaoa}
\end{equation}
where \(\gamma,\beta\) are the QAOA parameters, <jk> is an edge in the input graph, and \(E_{jk}\) has the form:
\begin{equation}
E_{<jk>} = U^{\dagger}(C, \gamma_1)...U^{\dagger}(B, \beta_p)C_{<jk>}U(B,\beta_p)...U(C,\gamma_1)
\label{eq:complex_form}
\end{equation}

The optimization process aims to find \(\gamma',\beta'\) such that \(E(\gamma',\beta')\) is minimized. This is done with gradient descent and is guided by the gradient \(\nabla E\), as shown in Equation~\ref{eq:sum_qaoa}:

\begin{equation}
\nabla E = \sum\limits_{<jk>} \nabla E_{<jk>}
\label{eq:sum_qaoa}
\end{equation}

In prior works~\cite{galda_transferability_2021, shaydulin_parameter_2023}, they identified that, if two energy functions, $E_1$ and $E_2$ have their gradient functions meet the criteria shown in Equation~\ref{eq:graph_same_optima}:

\begin{equation}
    \nabla E_1 = n * \nabla E_2
    \label{eq:graph_same_optima}
\end{equation}
then, these two functions share the same optima. Thus, the optimal parameters can be transferred between them. However, the likelihood of two random QAOA instances meeting the criteria in Equation \ref{eq:graph_same_optima} is extremely rare. 

\noindent \textbf{\ours}: The key insight in the design of \ours is that we can potentially approximate \(\nabla E\) by only a subset of terms \(E_{<jk>}\), if their gradient behaves similarly. That is:
\begin{equation}
\nabla E \approx \sum\limits_{<jk> \in S} \nabla E_{<jk>}
\label{eq:graph_approx}
\end{equation}
for some set \(S\subset{\text{all edges}}\). This would allow \ours to reduce the number of qubits and quantum gates required by eliminating local energy terms while adequately approximating the optimization landscape.

However, analytically determining which $E_{<jk>}$ terms can be eliminated is intractable due to their complex unitary structure, as shown in Equation \ref{eq:complex_form}. The entanglement between unitaries makes it challenging to rigorously model the impact of removing individual $E_{<jk>}$ terms. Therefore, \ours relies on an empirical node degree heuristic to select the dominant $E_{<jk>}$ terms. 

\subsection{Identifying Equivalent Instances}
\label{sec:equavalent_instances}
In Section~\ref{sec:transfer_theory}, we discussed that the overall expectation value of QAOA is the sum of individual sub-terms. Each sub-term corresponds to a subgraph encompassing all nodes and edges within a distance of $p$ from the central edge. This formulation leads to a generalized subgraph matching problem, where finding another graph with comparable subgraphs can be challenging due to the exponential search space.

The creation of each subgraph involves several steps. First, we select a main edge connected to the problem we are trying to solve, called the problem Hamiltonian. We then add nearby nodes and edges to form the subgraph. This expansion continues until a certain distance, denoted as $p$, is reached from the main edge. The number of nodes and edges added at each step depends on the degrees of the nodes already present in the subgraph. If two graphs exhibit similar average numbers of connections per node, called Average Node Degrees (AND), they likely possess identical subgraphs.

\begin{figure}[t]
    \centering
    \includegraphics[width=0.95\linewidth]{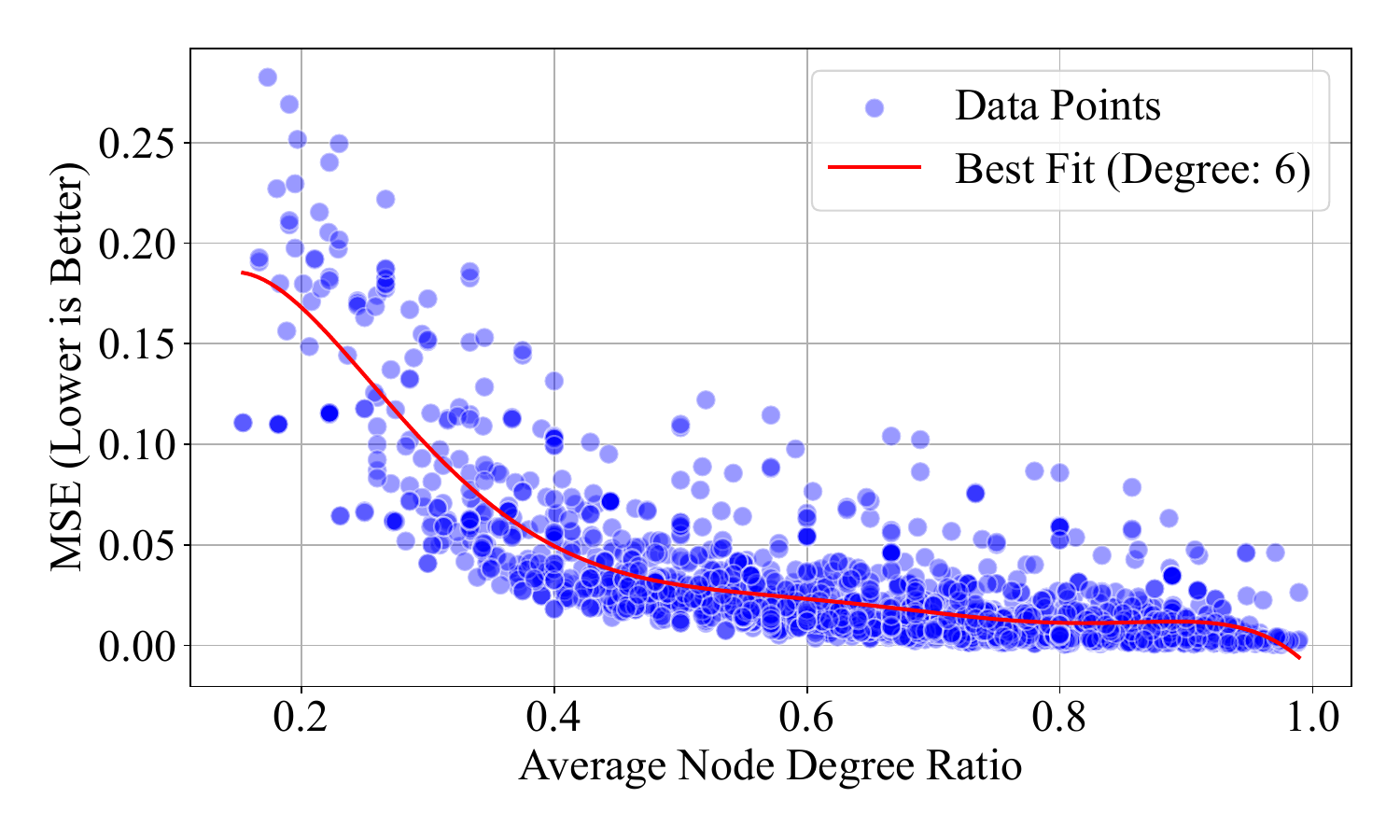}
    \caption{Scatter plot illustrating the correlation between the mean squared error (MSE) and the difference in Average Node Degrees (AND) of the subgraph with its original graph, the subgraphs are selected from a collection of all unique non-isomorphic subgraphs for 15 graphs. A 6th-degree polynomial was found to be the best-fit curse -- essentially indicating a correlation between MSE and AND.}
    \label{fig:mse_vs_and}
\end{figure}

To demonstrate this relation, we select 15 graphs at random and extract \emph{all of their} unique non-isomorphic subgraphs. We perform a 1-layer QAOA run for each subgraph using a grid search with a width of 30, resulting in 900 sets of parameters. We then normalize the expectation values and compute the mean squared error (MSE) between the subgraph results and its corresponding original graph. Figure~\ref{fig:mse_vs_and} shows the resulting plot, where the y-axis shows MSE and the x-axis shows the AND ratio, the proportion of the subgraph's AND to that of the original graph. The plot reveals a significant correlation between the variables, suggesting that smaller graphs with AND values comparable to the original graph can effectively identify optimal parameters.

\subsection{Mean Square Error for Landscape Similarity}
\begin{figure}[t]
    \centering
    \includegraphics[width=\linewidth]{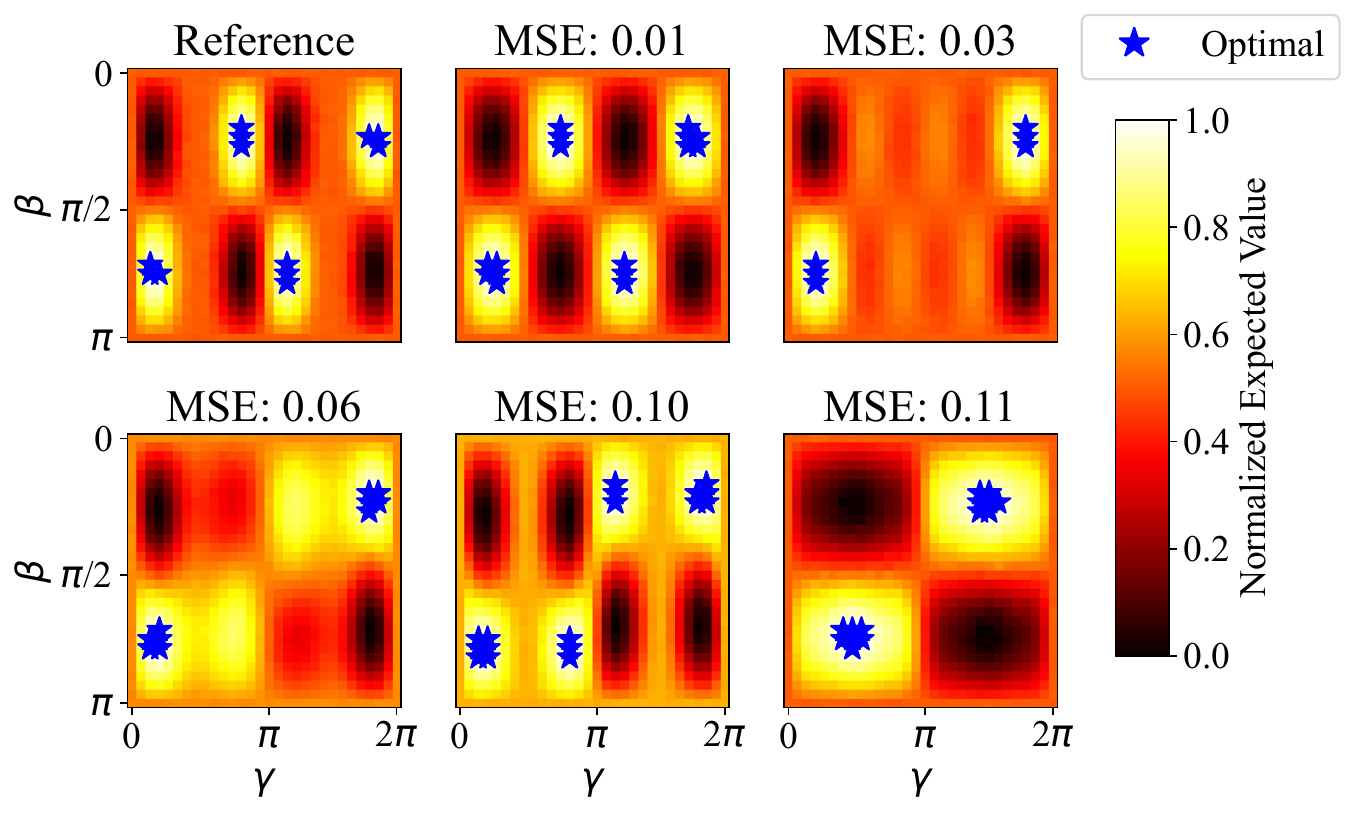}
    \caption{Six QAOA energy landscapes compared, with optimal points marked by blue stars. Mean Squared Error (MSE) values indicate similarity or divergence from the baseline landscape. A low MSE value, such as that of the 2nd graph in the first row (MSE = 0.01), indicates a landscape closer to the reference landscape. In contrast, a high MSE value, such as that of the graph in the bottom left (MSE = 0.11), indicates a landscape farther from the reference landscape. Our paper aims to identify an equivalent subgraph with an energy landscape that deviates less than 0.02 from the original graph.}
    \label{fig:design_mse}
\end{figure}

\begin{figure}[b]
    \centering
    \includegraphics[width=0.7\linewidth]{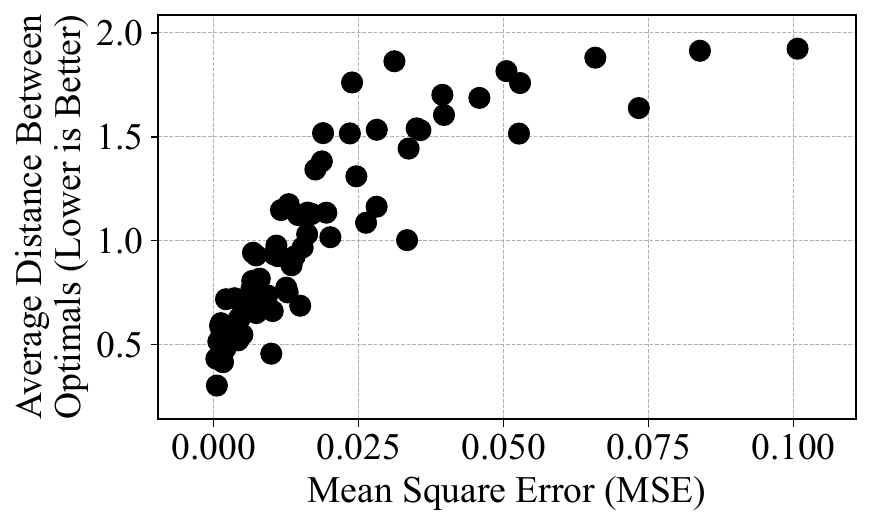}
    \caption{Scatter plot of MSE vs. average optimal point distance for random 15-node graphs and subgraphs, showing a strong correlation. This supports using MSE as a metric for energy landscape divergences.}
    \label{fig:mse_casestudy}
\end{figure}
A lower Mean Squared Error (MSE) implies higher similarity among instances during QAOA parameter optimization. However, defining an ideal MSE target remains challenging.

To establish an acceptable MSE, we analyze the optimal points' positions on energy landscapes. Figure~\ref{fig:design_mse} displays energy landscapes for six randomly selected graphs. One graph serves as a reference, and we calculate the MSE relative to this baseline for the remaining graphs, with optimal points denoted by blue stars. The normalized MSE can be viewed as a percentage error -- 0.01 MSE corresponds to a 1\% error.

Our observations suggest that when the MSE exceeds 2\% (0.02), optimal point placement significantly deviates from the reference landscape. In practice, we identify an equivalent subgraph with an energy landscape deviating less than 0.02 from the original graph. In Figure~\ref{fig:design_mse}, this 0.02 threshold equates to a minimum acceptable AND ratio of 0.7, and this ratio is used as the default value in the experiments. Users can adjust this threshold to suit their specific needs and use cases for equivalent instance searching.

To demonstrate the effectiveness of the MSE metric in quantifying differences among energy landscapes, we conducted a case study using random 15-node graphs and their subgraphs. Energy landscapes were generated for the original graph and its subgraphs using a 2-layer QAOA with 2048 random parameter sets. We computed the MSE between each subgraph's normalized landscape and that of the original graph. Additionally, we determined the average distance between their optimal solutions. Figure~\ref{fig:mse_casestudy} illustrates a strong correlation between MSE and the distance between optimal solutions. This correlation confirms that MSE accurately captures disparities in optimal solutions, making it a suitable metric for comparing energy landscapes in QAOA.

\subsection{Looking Beyond Pooling: Simulated Annealing}
Simulated annealing (SA) is a stochastic optimization algorithm inspired by metallurgical annealing~\cite{kirkpatrick_optimization_1983, laarhoven_simulated_1992}. SA begins with an initial solution and iteratively explores the solution space, accepting new solutions based on differences in objective function values and a decreasing 'temperature' parameter.

SA aligns well with our goal of identifying equivalent graphs. SA offers a more flexible and adaptive approach than graph pooling methods, which may impose rigid structures or rely on specific graph properties. This flexibility allows SA to discover subgraphs that better preserve the essential characteristics of the original graph, potentially making it a superior method. A key innovation of \ours is its utilization of SA to identify high-quality equivalent reduced graphs within the context of QAOA.

\begin{algorithm}
\caption{Simulated Annealing (SA) for Graph Reduction}
\label{alg:sa}
\begin{algorithmic}[1]
\Procedure{SA}{$G, k, T_0, \alpha, T_f, \textit{is\_adaptive}$}
    \State $AND_{G} \gets \textit{CalculateAND}(G)$
    \State $S \gets \textit{RandomSubgraph}(G, k)$
    \State $T \gets T_0$
    \While{$T > T_f$}
        \State $S_{\textit{neighbor}} \gets \textit{RandomNeighbor}(S, G)$
        \State $f_S \gets \textit{Objective}(S, AND_{G})$
        \State $f_{S_{\textit{neighbor}}} \gets \textit{Objective}(S_{\textit{neighbor}}, AND_{G})$
        \If{$f_{S_{\textit{neighbor}}} < f_S$}
            \State $S \gets S_{\textit{neighbor}}$
        \Else
            \State $p \gets \textit{Random}(0, 1)$
            \If{$p < \exp{(-(f_{S_{\textit{neighbor}}} - f_S) / T)}$}
                \State $S \gets S_{\textit{neighbor}}$
            \EndIf
        \EndIf
        \If{$\textit{is\_adaptive}$}
            \State $T \gets \alpha(T) * T$
        \Else
            \State $T \gets \alpha * T$
        \EndIf
    \EndWhile
    \State \Return $S$
\EndProcedure
\end{algorithmic}
\end{algorithm}

\ours utilizes an SA algorithm that supports `constant' and `adaptive' cooling schedules to dynamically construct reduced graphs and adjust the reduction ratio. The SA algorithm is executed multiple times. After each iteration, the average node degree (AND) of the resulting reduced graph is checked against the desired AND of the original graph. If the AND requirement is unmet, our algorithm adjusts the reduction ratio and reruns the SA algorithm until the desired AND is achieved. The resulting reduced graphs are then evaluated using the Mean Squared Error (MSE) metric to assess their similarity to the original graph.

Algorithm~\ref{alg:sa} shows the pseudocode of the proposed SA algorithm. The inputs to the algorithm include an input graph ($G$), the desired subgraph size ($k$), an initial temperature ($T_0$), a cooling rate function or factor called $\alpha$, a stopping temperature ($T_f$), and a boolean flag called $is\_adaptive$ that determines which cooling schedule to use. The algorithm starts by initializing a random subgraph as the initial solution and setting the initial temperature. Then, at each iteration, the algorithm explores a neighboring subgraph by replacing one of the nodes in the current subgraph with a node outside of it. The quality of the subgraph is measured using an objective function that calculates the difference between the ANDs of the subgraph and the original graph.

\begin{figure}[b]
    \centering
    \includegraphics[width=0.95\linewidth]{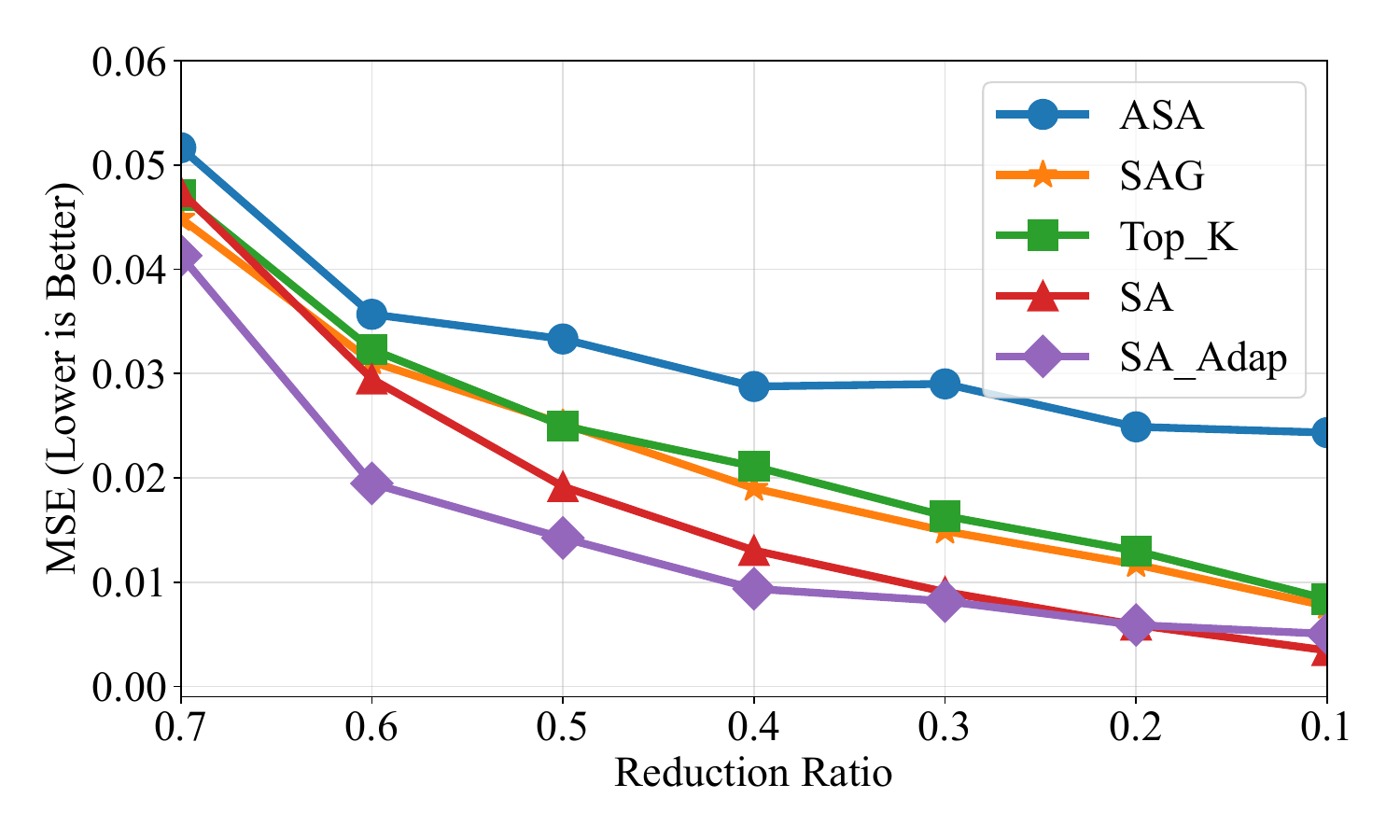}
    \caption{The mean square error (MSE) of subgraphs compared to the reduction ratio. We compare our Simulated Annealing (SA) methods to the state-of-the-art graph pooling techniques. The SA-based methods almost always provide lower MSE than prior techniques.}
    \label{fig:gnn_comp}
\end{figure}

\begin{figure*}
    \centering
    \includegraphics[width=0.9\linewidth]{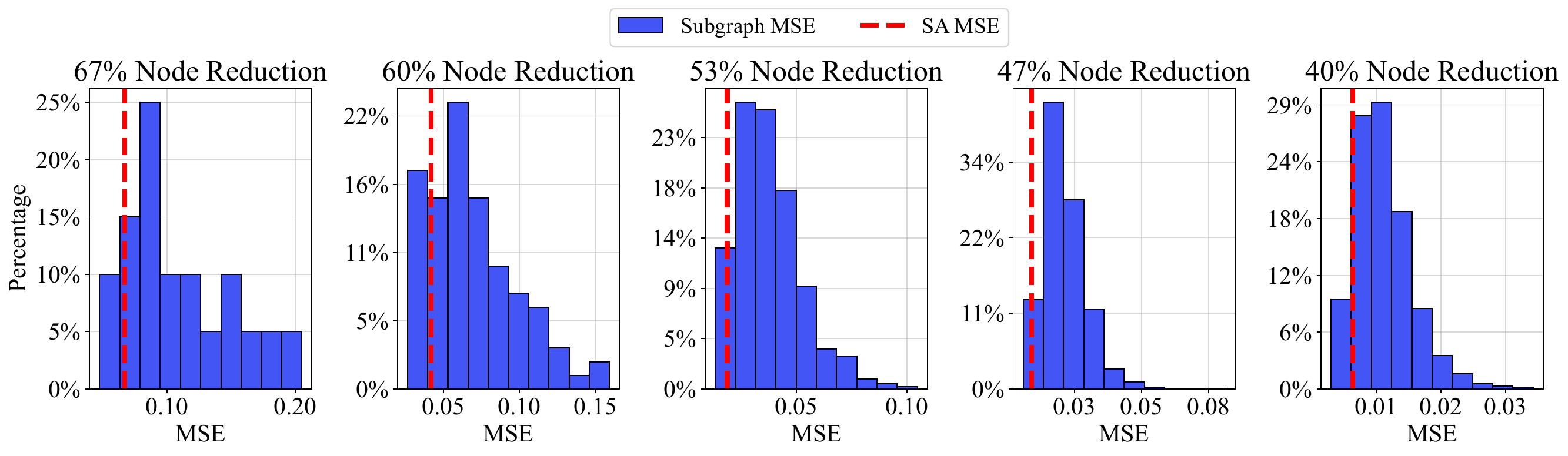}
    \caption{The performance of the proposed SA (simulated annealing) algorithm with varying node reduction ratios. The x-axis in each figure represents the mean square error, while the y-axis shows the percentage frequency. The algorithm, denoted as \ours, consistently identifies one of the most effective subgraphs across different reduction ratios. This suggests that \ours can consistently achieve desirable results for a given reduction ratio.}
    \label{fig:sa_performance}
\end{figure*}
The algorithm uses a temperature-dependent probability function to accept or reject neighboring subgraphs. If a neighboring subgraph has a better objective function value, then, it is accepted as the new solution. Otherwise, if it has a worse or equal objective function value, it is accepted with a probability that decreases as the temperature decreases. This probability function allows the algorithm to escape local optima in the early stages of the search while converging towards the global optimum as the temperature drops. It then updates the current temperature either by a constant factor or adaptively based on the number of rejected subgraphs.

The adaptive cooling schedule is a crucial component of the algorithm, as it controls the \emph{exploration}-\emph{exploitation} trade-off in the search process. By adjusting the cooling rate based on the current temperature, the algorithm can be fine-tuned to perform better on various graph pruning instances.

\subsection{SA versus Graph Pooling Methods}
\label{sec:pooling_compare}

We compare our SA-based methods to state-of-the-art graph neural network (GNN) pooling methods: Additive Self-Attention (ASA)~\cite{ranjan_asap_2020}, Set Attentional Aggregation (SAG)~\cite{knyazev_understanding_2019, lee_self-attention_2019}, and Top-k Pooling (Top\_k)~\cite{cangea_towards_2018, gao_graph_2019, knyazev_understanding_2019}. These pooling methods use fixed reduction ratios and do not dynamically check if the reduced graph accurately approximates the original graph. We test on the random graph dataset with $p=3$, using a range of fixed reduction ratios from 0.1 to 0.7.

Figure~\ref{fig:gnn_comp} shows the experiment's results, including both `constant' and `adaptive' cooling versions. The results show that both these versions outperform all state-of-the-art GNN-based graph pooling methods, except for constant cooling when the reduction ratio is 0.7. However, this reduction ratio is too extreme and impractical in real-world applications. Overall, the adaptive cooling version of SA performs significantly better than the other methods. Given that adaptive cooling has a lower computational overhead, we equip \ours to employ the adaptive cooling method in all cases.

The superior performance of \ours compared to other GNN pooling methods highlights the importance of developing specialized graph reduction techniques tailored to the needs and constraints of QAOA and quantum computing.

\subsection{Effectiveness of Simulated Annealing}
To demonstrate the effectiveness of our proposed algorithm, we conducted an experiment using a random 15-node graph. We analyze its unique connected subgraphs for node reduction ratios of 0.67, 0.53, and 0.40, and for each subgraph, we perform a grid search with 900 data points and calculate the normalized MSE. Figure~\ref{fig:sa_performance} shows the results as a histogram, with the x-axis representing MSE values and the y-axis indicating the frequency of subgraphs as a percentage. A dashed red line is marked to show the MSE obtained using the proposed SA algorithm. Across all reduction ratios, the SA algorithm consistently achieves the lowest MSEs.

\section{Methodology}

\subsection{Figure of Merit}
\label{sec:merit}

We utilize two key performance metrics - Mean Square Error (MSE) and Approximation Ratio - to evaluate \ours. The MSE is primarily used to measure the similarity between two QAOA instances, capturing discrepancies in the energy landscapes. We apply MSE in two settings: ideal execution to compare the energy landscapes of the baseline graph and \ours graph, and noisy execution to compare both the noisy baseline and \ours landscapes against the ideal baseline. On the other hand, the Approximation Ratio is used to evaluate the quality of QAOA outcomes relative to the ground truth. This ratio assesses the performance of \ours against the baseline in ideal and noisy conditions. 

\subsubsection{Mean Square Error (MSE)}
Our study primarily uses the MSE to measure the similarity between two QAOA instances. MSE serves as a tool to quantify how closely two different QAOA-generated energy landscapes resemble each other. The MSE is defined by the equation:

\begin{equation}
    MSE = \frac{1}{N}\sum_{i=1}^{N}(E_i - \hat{E}_i)^2
    \label{eq:mse}
\end{equation}
where $N$ is the total number of data points, the default is set to 1024 in our experiments. $E_i$ represents the normalized energy at the $i^{th}$ data point in the first QAOA instance, and $\hat{E}_i$ corresponds to the normalized energy at the same point in the second QAOA instance. This formula effectively captures the discrepancy between the two energy landscapes, providing a numerical measure of their similarity.

Our application of MSE occurs in two distinct settings. In the ideal execution setup, MSE is employed to compare the energy landscapes generated by the baseline graph and \ours's graph. The aim is to assess how closely the \ours graph's landscape mirrors the baseline under ideal conditions. In the noisy execution setup, on the other hand, two separate MSE values are computed. The first is between the noisy baseline landscape and the ideal baseline landscape. The second MSE is between the noisy \ours landscape and the ideal baseline landscape. In this context, the ideal baseline landscape serves as a benchmark, and our goal is to demonstrate that the \ours graph under noisy conditions can produce a landscape more akin to this ideal baseline than what is achieved with the noisy baseline.

\subsubsection{Approximation Ratio.}
In addition to MSE, we utilize the approximation ratio $r$ to evaluate the quality of QAOA outcomes. This ratio is defined as the ratio of the optimal expectation value obtained by QAOA to the ground truth result, determined classically via brute force:

\begin{equation}
    r = \frac{\min_{\boldsymbol{\gamma}, \boldsymbol{\beta}} \langle \psi(\boldsymbol{\gamma}, \boldsymbol{\beta}) | C | \psi(\boldsymbol{\gamma}, \boldsymbol{\beta}) \rangle}{C_{\text{ground truth}}}
\end{equation}
where $|\psi(\boldsymbol{\gamma}, \boldsymbol{\beta}) \rangle$ represents the trial state prepared by the QAOA operator given in Equation~\ref{eq:trail_state}. Our experiments aim to compare the approximation ratios obtained from the baseline and \ours in both ideal and noisy execution setups. In the ideal execution setup, we seek approximation ratios for \ours that closely match the baseline. On the other hand, in the noisy execution setup, our objective is to assess the improvement brought by \ours in approximation ratio over the noisy baseline.

\subsection{Benchmark Graph Datasets}

To comprehensively evaluate \ours, we selected four diverse benchmark graph datasets: AIDS, Linux, IMDb, and a collection of Random graphs spanning different domains.

\begin{itemize}
\renewcommand{\labelitemi}{\scriptsize$\blacksquare$}
\item \textbf{AIDS:} A set of 700 chemical compound graphs from the National Cancer Institute's repository. Each graph represents a chemical compound. Specifically, the nodes of the graph are the atoms and its edges are chemical bonds. The average graph size is eight nodes.

\item \textbf{Linux:} Comprising 1,000 function call graphs extracted from the Linux kernel source code. Nodes represent functions, and edges denote function calls. The average graph size is ten nodes.

\item \textbf{IMDb:} Consisting of 1,500 movie collaboration networks from the Internet Movie Database. Nodes represent actors, and edges indicate collaborations between them. The average graph size is six nodes, with most containing fewer than ten nodes.

\item \textbf{Random graphs:} We generated ten random graphs using the NetworkX random graph generator, with node counts ranging from 7 to 20. These graphs provide a versatile testing platform for our experiments.
\end{itemize}

This diverse dataset selection allows us to evaluate our methods across various domains, offering insights into their scalability and adaptability to different graph structures. Table~\ref{tab:benchmark_dataset} summarizes the characteristics of benchmark graph datasets used in our experiments.

\begin{table}[b]
\centering
\small
\caption{\label{tab:benchmark_dataset} Description of Benchmark Graph Datasets}
\begin{tabularx}{\columnwidth}{>{\centering\arraybackslash}X >{\centering\arraybackslash}X >{\centering\arraybackslash}X >{\centering\arraybackslash}X} 
\toprule
\textbf{Dataset} & \textbf{Description} & \textbf{Number of Graphs} & \textbf{Number of Nodes} \\ 
\midrule
AIDS~\cite{da_vitoria_lobo_iam_2008} & Chemical Compounds & 700 & 2-10 \\ 
\midrule
LINUX~\cite{wang_efficient_2012} & Program Dependence & 1000 & 4-10 \\ 
\midrule
IMDb~\cite{yanardag_deep_2015} & Ego Networks & 1500 & 7-89  \\
\midrule
Random~\cite{erdHos1960evolution} & Erdős-Rényi & 10 & 7-20  \\
\bottomrule
\end{tabularx}
\end{table}
\subsection{Circuit Simulation and Noise Modeling}
We use the Qiskit~\cite{Qiskit} framework for circuit simulations. Ideal quantum circuit simulations are executed with the statevector backend, while noisy simulations are performed with the density matrix backend to account for potential noise impact on quantum circuits. We employ the FakeToronto backend, which emulates the noise characteristics of the IBM Quantum Toronto device, incorporating gate errors, readout errors, and relaxation times. FakeToronto provides realistic quantum hardware conditions. Similar to prior work~\cite{dasjigsaw}, we transpile circuits using the SABRE algorithm~\cite{li2019tackling} to optimize depth and execution time. It selects the circuit with the shortest depth out of 100 repetitions. This process helps ensure optimal performance and resource usage.

\subsection{Hardware Platform}
Perlmutter's GPU nodes, each equipped with four Nvidia A100 GPUs (40GB VRAM per GPU), are used for efficient quantum circuit simulations, reducing the time required to obtain results. For executing quantum circuits on real hardware, we utilize the 27-qubit \emph{ibmq\_kolkata} backend from IBM and the larger 79-qubit \emph{Aspen-M-3} from Rigetti. These backends allow us to validate our simulation results in real-world settings, providing insights into circuit performance under hardware constraints and error rates. By comparing the outcomes of our simulations with the results from the device backends, we can assess the accuracy and robustness of our methodology in the presence of hardware imperfections.

\subsection{Graph Pooling Methods}
We compared our proposed method, which uses dynamic checking to ensure accurate graph reduction, with three fixed-ratio graph pooling methods: Additive Self-Attention (ASA)~\cite{ranjan_asap_2020}, Set Attentional Aggregation (SAG)~\cite{knyazev_understanding_2019, lee_self-attention_2019}, and Top-k Pooling (Top\_k)~\cite{cangea_towards_2018, gao_graph_2019, knyazev_understanding_2019}. These methods were chosen because they are widely used in the literature and have been shown to achieve state-of-the-art performance in various graph-related tasks~\cite{Fey/Lenssen/2019}.

All the graph pooling methods take the graph feature vector and a pooling ratio as inputs. In our case, the feature vector is generated from the input graph, which is a normalized vector that includes the node degrees, clustering coefficient, betweenness centrality, closeness centrality, and eigenvector centrality. These metrics provide insights into the node's connectivity, position within the network, and influence. The pooling ratio is determined dynamically based on the \ours reduced graph. We first generate a reduced graph using the \ours method. Then, we calculate the pooling ratio such that all pooled graphs are the same size as the \ours graph. By ensuring all reduced/pooled graphs are the same size, we enable a fair comparison between them.

\subsection{Parameter Transfer}
Previous studies on transferring optimal QAOA parameters focused on random regular graphs~\cite{shaydulin_parameter_2023}. To evaluate parameter transferability on non-regular graphs, we start with random regular base graphs. We then randomly modify a small portion, 10\% in our case, of the edges in these base graphs by removing some edges and adding new ones. This process makes the graphs slightly irregular while retaining similarities to the original regular base graphs. We generate two graphs to test parameter transfer: 1) a \ours graph, and 2) a smaller random regular graph with the same node degrees as the original unmodified base graph and the same number of nodes as the \ours graph. Comparing these two graphs allows us to evaluate how \ours performs on non-regular graphs compared to transferring parameters directly between similar non-regular graphs.

\section{Results}
\begin{figure}[b]
    \centering
    \includegraphics[width=0.8\linewidth]{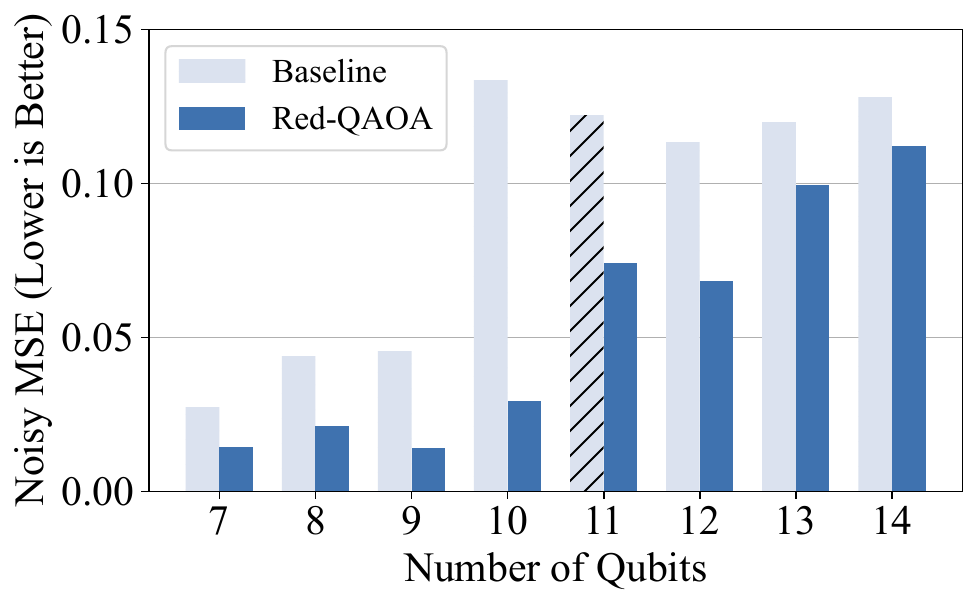}
    \caption{Mean Squared Error (MSE) comparison of baseline and \ours noisy landscapes for graphs of 7-14 nodes. As the number of qubits increases, the noisy MSEs for both \ours and the baseline increase due to a higher device noise impact. Overall, with a more noise-tolerant circuit, \ours consistently performs better than the baseline.}
    \label{fig:noisy_mse}
\end{figure}

\subsection{Effectiveness of Recovering the Ideal Landscape}

Our initial investigation assesses Red-QAOA's ability to restore the energy landscape distorted by noise. We compare the noisy MSE of the \ours landscape with the ideal QAOA landscape and contrast it with the noisy MSE between the baseline noisy QAOA landscape and the ideal QAOA landscape. Noisy MSE values evaluate energy landscape preservation by comparing noisy landscapes to ideal ones. Previously, we focused on ideal MSEs between different QAOA instances, capturing their equivalency. Therefore, the noisy and ideal MSEs should be interpreted separately.

We generate random graphs with 7 to 14 nodes and conduct noisy simulations. Figure~\ref{fig:noisy_mse} illustrates that \ours consistently outperforms the baseline noisy QAOA landscape in all scenarios. This improvement is primarily attributed to \ours' use of a reduced graph, reducing node counts by an average of 36\% and edge counts by an average of 50\%, significantly reducing the likelihood of noise interference.

\begin{figure}[t]
    \centering
    \includegraphics[width=\linewidth]{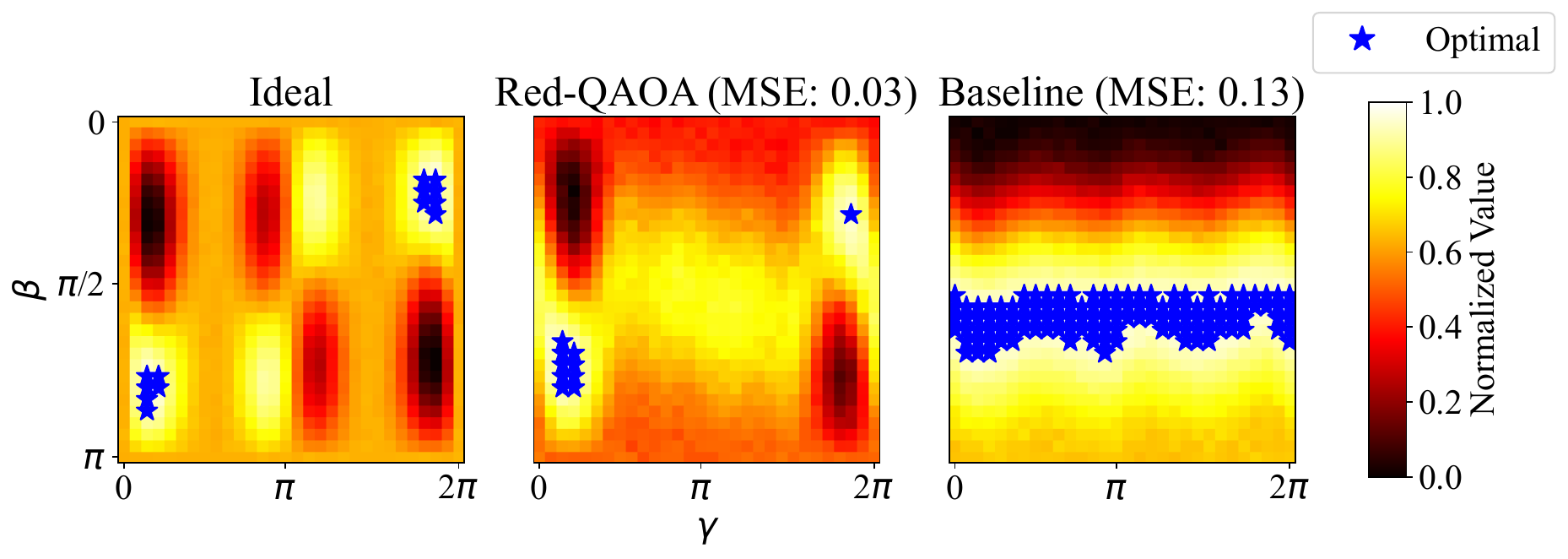}
    \caption{Normalized energy landscapes of ideal, noisy baseline, and \ours for the best-case scenario of the 10-node graph. Blue stars indicate the globally optimal points on each landscape. \ours outperforms the noisy baseline by locating optimal points closer to the ideal scenario.}
    \label{fig:noisy_land_10}
\end{figure}

Among the tested graphs, the 10-node graph exhibits the most substantial reduction in MSE from the baseline to \ours. Figure~\ref{fig:noisy_land_10} displays the energy landscapes for the ideal, \ours, and noisy baseline, with blue stars marking the globally optimal points. The baseline noisy landscape is severely affected by noise, while the \ours energy landscape closely retains the ideal landscape's overall shape. Importantly, the location of optimal points is crucial, representing final solutions. The baseline's noise introduces false global points, whereas \ours maintains optimal points very close to the ideal landscape.

\begin{figure}[b]
    \centering
    \includegraphics[width=\linewidth]{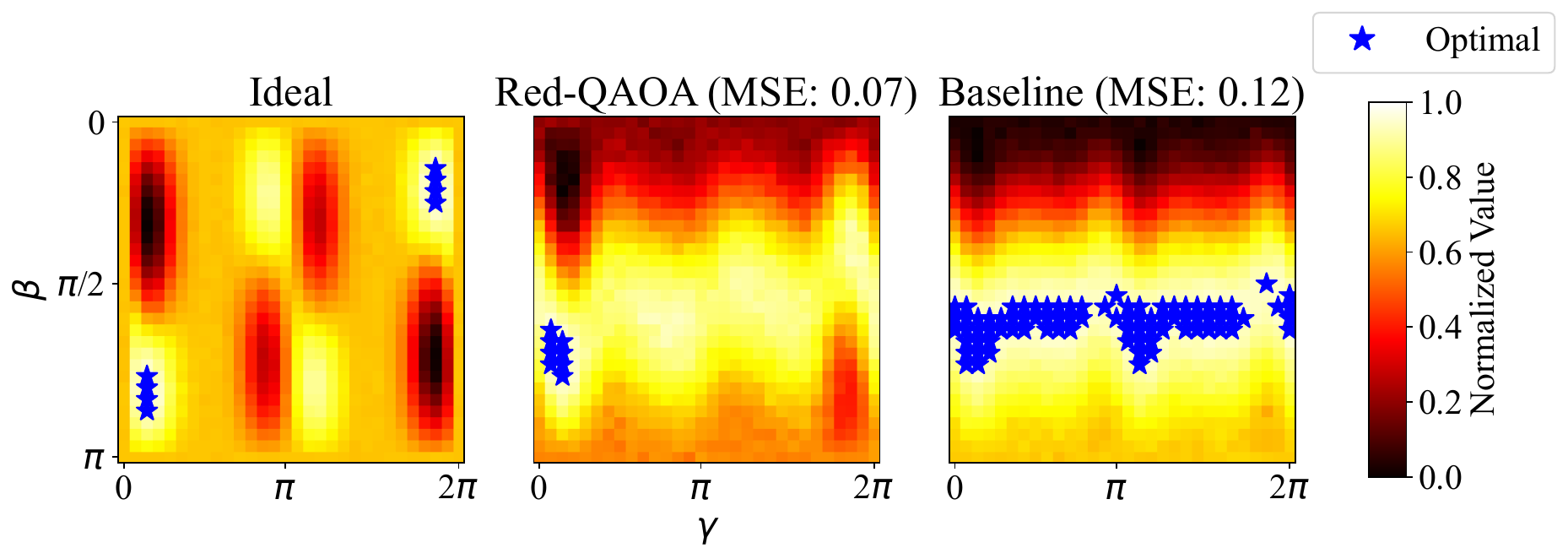}
    \caption{Normalized energy landscapes of ideal, noisy baseline, and \ours for the worst-case scenario of the 11-node graph. Blue stars indicate the globally optimal points on each landscape. The optimal points on \ours landscape begin to deviate from the ideal scenario; however, even with this deviation, \ours still outperforms the baseline, which exhibits a much greater deviation in terms of the location of optimal points from the ideal scenario.}
    \label{fig:noisy_land_11}
\end{figure}
In contrast, Figure~\ref{fig:noisy_land_11} illustrates a worst-case scenario for 11 nodes where we observe the slightest reduction in MSE compared to the baseline. In this case, \ours landscape's overall shape is noticeably off from the ideal case. However, its globally optimal points are still very close to the true optimal, indicating the superior performance of \ours. Interestingly, the ideal landscapes of the 10-node and 11-node graphs are quite similar since they have very close node and edge counts. This implies that the reduced graph found by \ours for the 10-node case can be used for the 11-node case to achieve a better result. However, due to the constraints on searching for subgraphs, the reduced graph was rejected by \ours. This indicates a potential opportunity for further improvements. 

Overall, with a smaller circuit, \ours is more robust to noise than the baseline approach and can perform better in practical quantum computing implementations.

\subsection{Reductions and Ideal MSEs on Small Graphs}

\begin{figure}[b]
    \centering
    \includegraphics[width=0.8\linewidth]{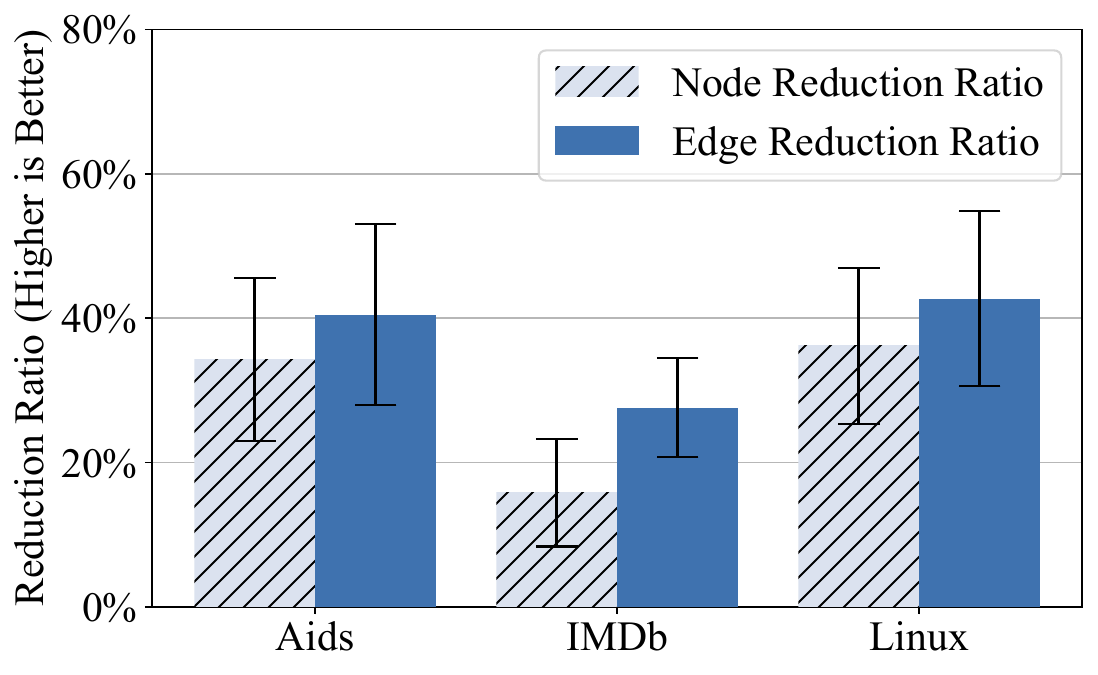}
    \caption{Graph reductions, measured in terms of node and edge reduction ratios, for graphs from Aids, Linux, and IMDb graph datasets with up to 10 nodes. On average, \ours achieves 28\% node reductions and 37\% edge reductions.}
    \label{fig:reductions}
\end{figure}

To thoroughly assess \ours performance, we selected real-world graphs with up to 10 nodes from Aids, Linux, and IMDb datasets for graph reduction and MSE experiments. We analyze larger IMDb graphs in Section~\ref{sec:medium_graphs} to evaluate \ours scalability and effectiveness. Figure~\ref{fig:reductions} shows the three graph datasets' node and edge reduction ratios. Each dataset is represented by two bars, one for the node reduction ratio and one for the edge reduction ratio. On average, 28\% of nodes were reduced, and 37\% of edges were eliminated, resulting in substantially smaller graphs for execution.

\begin{figure}[t]
    \centering
    \includegraphics[width=0.8\linewidth]{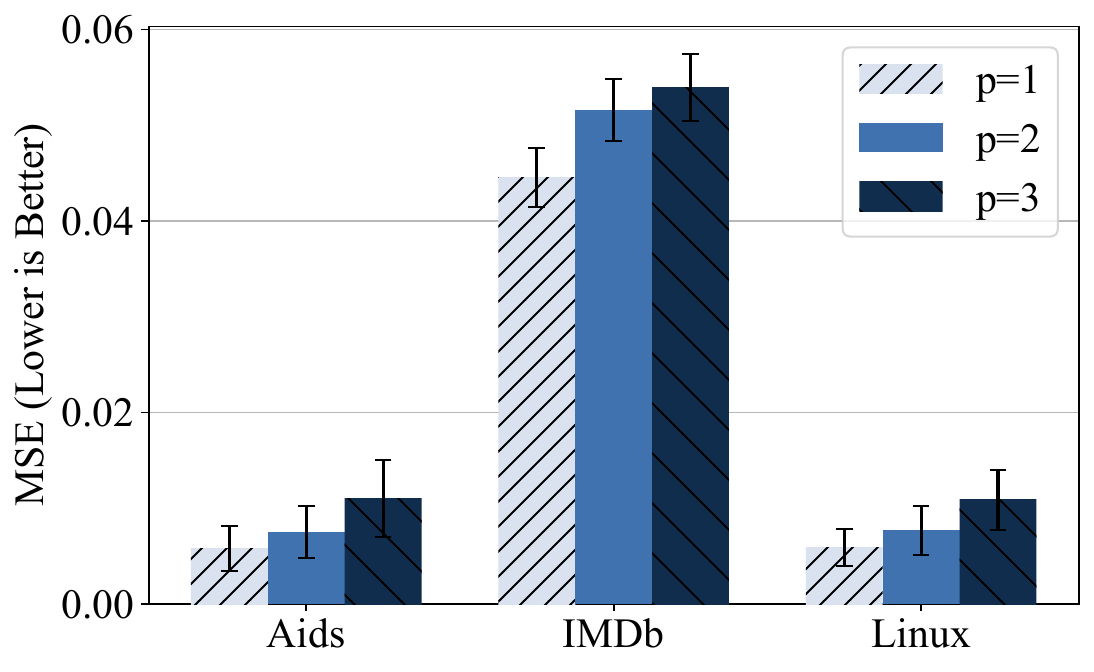}
    \caption{Mean Squared Error (MSE) for Aids, Linux, and IMDb Graph Datasets with Up to 10 Nodes for $p=1$, $p=2$, and $p=3$. The MSE achieved with the Aids and Linux datasets is below 0.01, while for IMDb, it is around 0.05. Section~\ref{sec:medium_graphs} presents more detailed information about this. }
    \label{fig:mses}
\end{figure}

Figure~\ref{fig:mses} displays the Mean Squared Error (MSE) for the three graph datasets, plotting results for different QAOA circuit layer parameter values: $p=1$, $p=2$, and $p=3$. For each $p$ value, we randomly selected 1024 parameter sets and computed the MSE relative to the baseline. Each dataset is represented in the figure by three bars corresponding to the distinct $p$ values. The data indicates that as $p$ increases, the MSE experiences a slight increase, suggesting that adding more layers to the QAOA circuit introduces complexity into the graph reduction process. However, this growth rate remains relatively low and manageable.

\begin{figure}[b]
    \centering
    \includegraphics[width=0.68\linewidth]{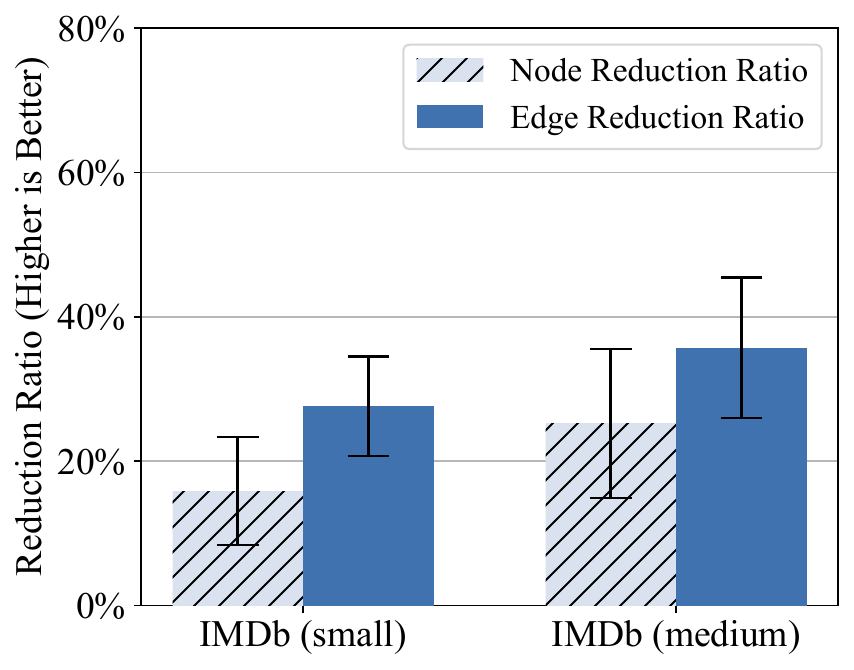}
    \caption{Node and edge reduction ratios for IMDb, classified into small (up to 10 nodes) and medium (up to 20 nodes) graph categories. When scaling from small to medium graphs, the node reduction ratio increased from 15\% to 25\% and the edge reduction ratio rose from 28\% to 35\%.}
    \label{fig:reductions_imdb}
\end{figure}

\subsection{Scaling Up to Larger Graphs}
\label{sec:medium_graphs}

Figure~\ref{fig:reductions} and Figure~\ref{fig:mses} reveal a significant observation: the IMDb dataset has the lowest reduction ratios among the three datasets and its MSE values are notably higher. This difference primarily stems from IMDb's considerably higher average node degrees than AIDS and Linux datasets. Since our analysis focuses on graphs with a maximum of 10 nodes, removing a single node from IMDb's graphs results in a larger loss of edges. This discrepancy is evident in Figure~\ref{fig:reductions}, where the gap between edge and node reduction ratios is about 5\% for AIDS and Linux but exceeds 10\% for IMDb.

\begin{figure}[t]
    \centering
    \includegraphics[width=0.68\linewidth]{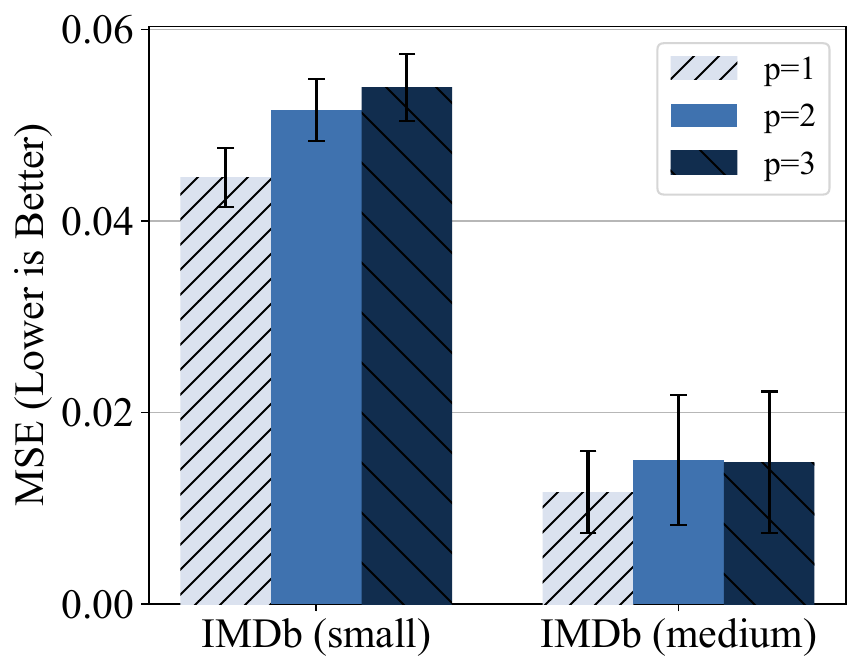}
    \caption{Mean Squared Error (MSE) for IMDb datasets for $p=1$, $p=2$, and $p=3$. Graphs are divided into small (up to 10 nodes) and medium (up to 20 nodes) graphs. We observed a notable reduction in overall MSEs, dropping from approximately 0.05 to below 0.02.}
    \label{fig:mses_imdb}
\end{figure}

Figure~\ref{fig:reductions_imdb} and Figure~\ref{fig:mses_imdb} show the results for IMDb graphs in two categories: IMDb small (up to 10 nodes) and IMDb medium (10 to 20 nodes). As graph size increases, the reduction ratio improves, and MSEs decrease. In this case, the IMDb medium graph set exhibits a similar performance as compared to Aids and Linux datasets. This suggests that \ours has a relatively worse performance for small, dense graphs. However, for these graphs, the noise impact and execution overhead are considerably smaller than in larger graphs, making it beyond the scope of Red-QAOA.

\subsection{Scalability of \ours}

\begin{figure}[b]
    \centering
    \includegraphics[width=0.9\linewidth]{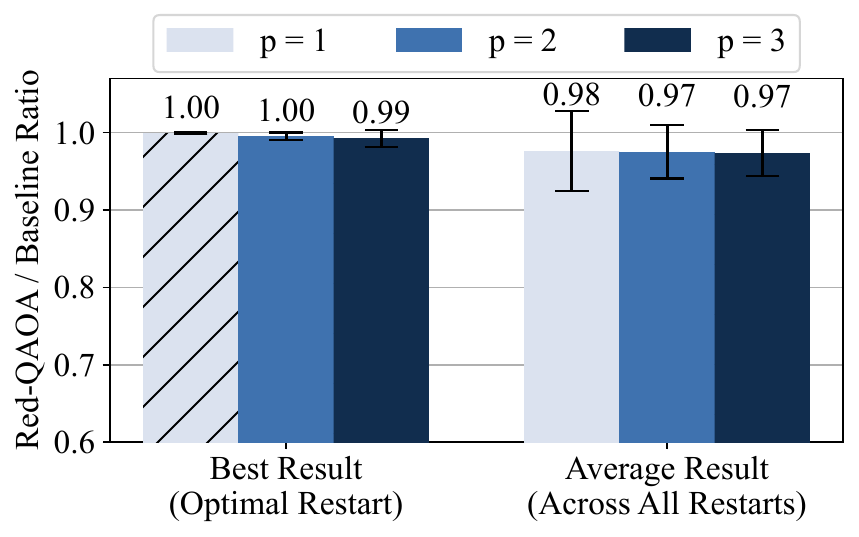}
    \caption{\ours achieves near-optimal solutions for all testing cases when considering the best results achieved. On average, \ours achieves over 97\% performance relative to baseline QAOA on a set of 100 large 30-node graphs.}
    \label{fig:scability_effectiveness}
\end{figure}
\subsubsection{Effectiveness}
To assess the scalability of \ours, we conducted tests on 100 randomly generated graphs, each containing 30 nodes. We utilized the COBYLA classical optimizer with 20, 50, and 150 restarts for 1, 2, and 3 QAOA layers, respectively. Two key performance metrics were evaluated: the best result among all restarts for a given layer and the average result across restarts. Figure \ref{fig:scability_effectiveness} demonstrates that \ours consistently achieved near-optimal best results, exceeding 99\% across all cases compared to the baseline. Remarkably, even with an average reduction of 30.7\% in nodes and 44.3\% in edges, \ours maintained over 97\% of the baseline's average performance.

\subsubsection{Runtime Analysis}

\begin{figure}[h!]
    \centering
    \includegraphics[width=\linewidth]{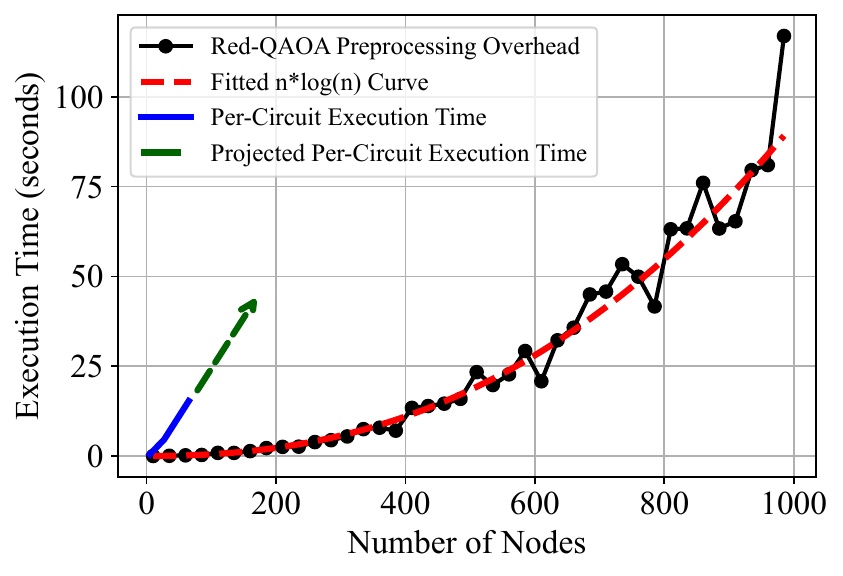}
    \caption{The runtime overhead of \ours scales as O($n\log n$), showing asymptotic efficiency. Using limited benchmark data from~\cite{wack2021quality}, we extrapolated the per-circuit execution time up to 65 qubits and compared \ours against it. Overall, \ours introduces a \textit{negligible} graph processing overhead as compared to circuit execution time.}
    \label{fig:runtime}
\end{figure}

\ours imposes minimal preprocessing overhead, scaling as \texttt{$n\log n$} due to its binary search approach over graph sizes. Figure \ref{fig:runtime} confirms this asymptotically efficient complexity on random graphs ranging from 10 to 1000 nodes. For a small 10-node graph, \ours requires just 0.004 seconds of preprocessing time. In contrast, executing a corresponding 1-layer QAOA circuit on ibm\_sherbrooke processor takes 4.2 seconds. Therefore, \ours has negligible overhead, around 0.1\% of the total QAOA runtime, even for modest problem sizes. Theoretical \texttt{$n\log n$} scaling suggests \ours will maintain efficient preprocessing as problem dimensions and quantum system sizes increase. This negligible overhead makes \ours well-suited for time-critical applications.

\subsection{End to End Evaluation}
To evaluate graph pooling methods for generating surrogate problems to train QAOA, we tested four techniques: \ours, SAG, Top-K, and ASAPooling. We generated 10-node random graphs and reduced versions using each method.

We performed grid searches to find the optimal QAOA parameters on the original and surrogate graphs. The key metric is the relative improvement in the approximation ratio over the original graphs under the noisy execution setup.

\begin{figure}[h]
    \centering
    \includegraphics[width=0.8\linewidth]{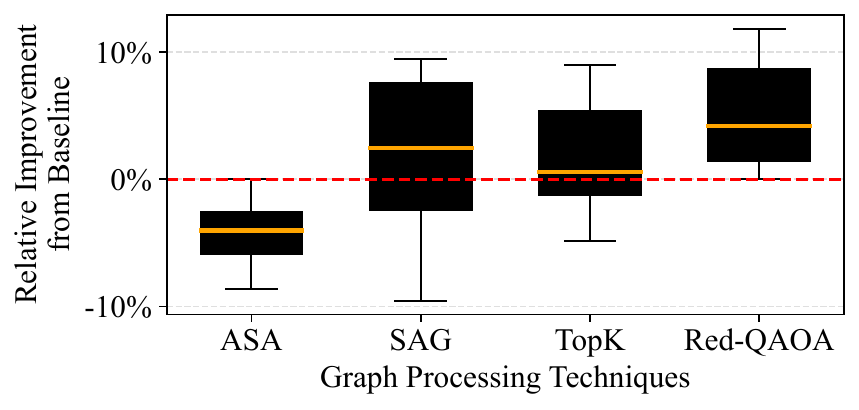}
    \caption{Box plots of relative improvement in approximation ratio over noisy baselines for \ours and three GNN-based graph pooling techniques. \ours showed consistently positive improvements across all testing cases, outperforming the highly variable SAG and Top-K pooling. The ASAPooling performs the worst overall.}
    \label{fig:end_to_end}
\end{figure}

Figure~\ref{fig:end_to_end} shows that \ours provides consistent positive improvements, with a 4.2\% median increase over the baseline. This outperformed the other techniques, with SAG and Top-K pooling showing high variability. ASAPooling consistently decreased performance.

\begin{figure}[h]
    \centering
    \includegraphics[width=0.85\linewidth]{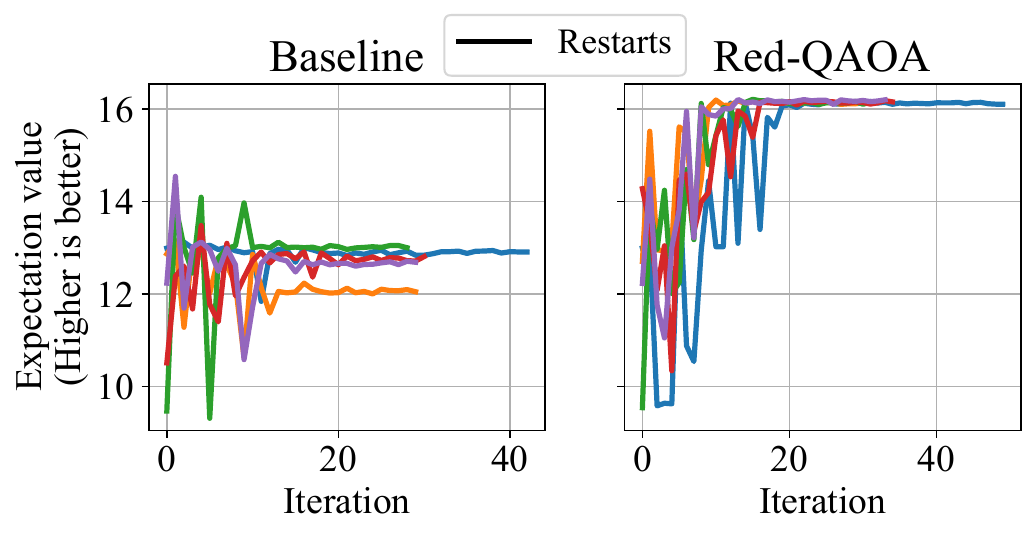}
    \caption{Comparison of convergence between standard QAOA and \ours on noisy quantum simulations. Five restarts are performed using the baseline and reduced graphs from \ours. \ours demonstrates substantially faster and better convergence to optimal-energy solutions compared to the standard QAOA optimization.}
    \label{fig:end_to_end_timeseries}
\end{figure}

To evaluate the convergence rate improvement of QAOA using \ours, a 10-node random graph was selected, and classical optimization with five random restarts was performed using the COBYLA optimizer~\cite{powell1994direct}. Figure \ref{fig:end_to_end_timeseries} illustrates the convergence behaviour of noisy QAOA simulations on this problem instance using both the original graph and the \ours graph. To enable a direct comparison of convergence rates, the QAOA parameters were recorded at each iteration and then used to re-calculate the expected energy with an ideal noiseless simulator. Overall, \ours shows substantially faster and better convergence to high-energy solutions than the standard QAOA optimization.

\subsection{Comparison: Parameter Transfer}

\begin{figure}[t]
    \centering
    \includegraphics[width=0.8\linewidth]{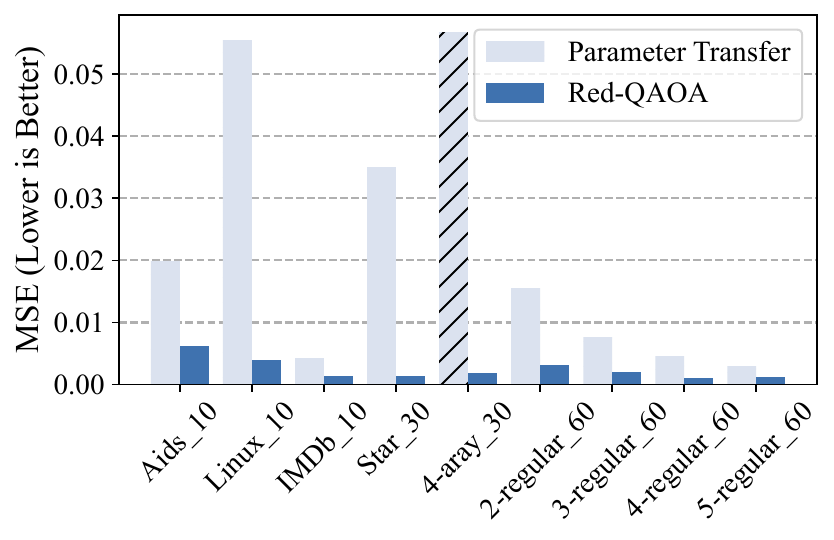}
    \caption{MSE between ideal and transferred QAOA landscapes using \ours and parameter transfer, evaluated on real-world and non-regular graphs. \ours reliably outperforms transfer across graph types.}
    \label{fig:larger_param_transfer}
\end{figure}

Previous research~\cite{galda_transferability_2021} demonstrated QAOA parameter transferability on random \emph{regular} graphs. Our experiments evaluated this transferability on various graphs up to 60 nodes, including real-world and non-regular graphs from AIDS, Linux, and IMDb datasets. We also tested modified regular and non-regular star/4-array graphs. We transferred optimal parameters between graphs with even/odd degree nodes for parameter transfer. To ensure fairness, we initially reduced the graph using \ours and then created a random regular graph with a similar node count and average degree. Figure \ref{fig:larger_param_transfer} displays the MSE between ideal and transferred landscapes for both methods. Parameter transfer works well for regular or near-regular graphs but struggles with increased randomness. In contrast, \ours consistently maintains a low MSE across all graph types, demonstrating robust performance regardless of regularity.

\subsection{Execution on Real Quantum Devices}

\begin{figure}[b]
    \centering
    \includegraphics[width=0.98\linewidth]{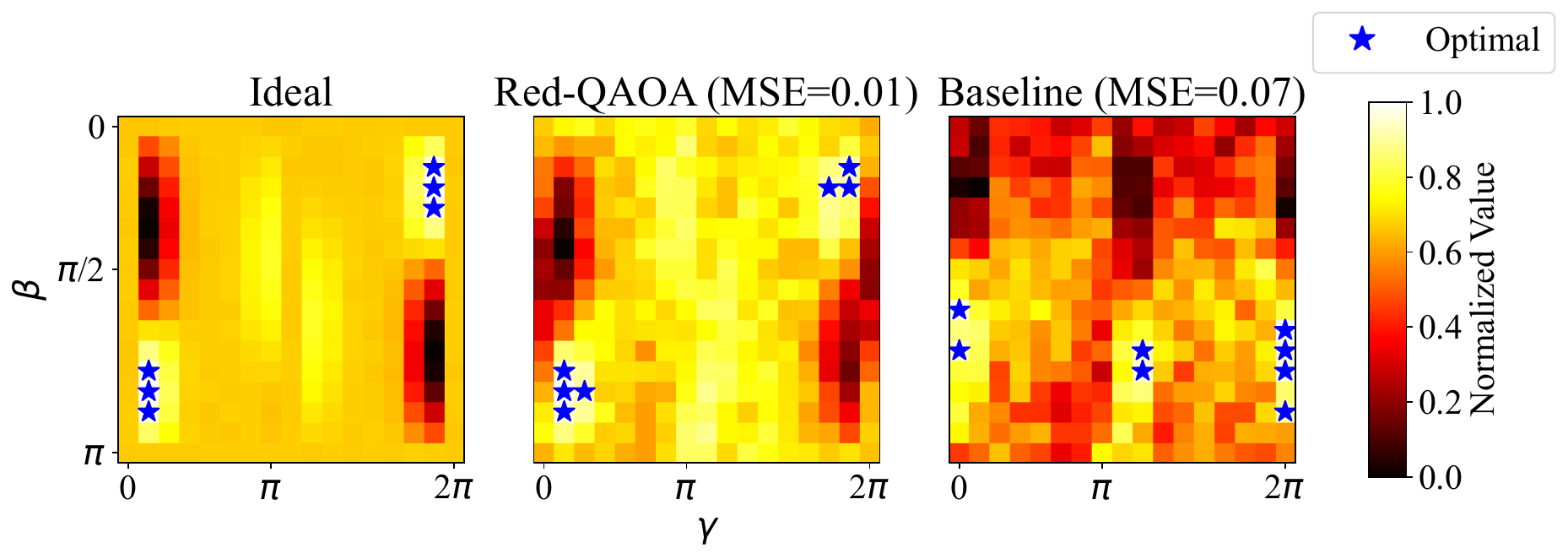}
    \caption{Normalized energy landscapes of an ideal scenario, a noisy baseline, and \ours for a random 13-node graph executed on the 27-qubit ibmq\_kolkata backend are compared. The optimal points on the \ours energy landscape are much closer to the ideal case, whereas the baseline shows a significant deviation.}
    \label{fig:device_land}
\end{figure}

\noindent \textbf{IBM Device}: We demonstrate Red-QAOA's effectiveness in addressing noise and hardware limitations on a real quantum device. We employ QAOA on the 27-qubit ibmq\_kolkata backend with a 13-node random graph. Scaling to larger graphs is currently challenging due to the significant device error rate. Though we have not extensively optimized the circuit beyond transpile the circuit using SABRE~\cite{li2019tackling} multiple times and selecting the minimum-depth one. Further optimizations~\cite{alam2020circuit, ayanzadeh2023frozenqubits} and error mitigation strategies~\cite{temme2017error} can potentially reduce noise impact for larger circuits. However, our primary aim is to showcase the improvement achieved with \ours over noisy baseline under the same execution conditions. This enhancement is expected to persist for larger instances and with more rigorous optimizations.

Figure~\ref{fig:device_land} presents normalized energy landscapes for the ideal scenario, a noisy baseline, and \ours. This set of results showcases \ours's accuracy and reliability in identifying equivalent instances and reducing noise impact in QAOA parameter optimization. \ours outperforms the noisy baseline in identifying optimal points in the energy landscape, substantiating its effectiveness in addressing noise and hardware constraints during QAOA parameter optimization. This noise reduction leads to more accurate and reliable energy landscapes, resulting in more efficient solutions for large-scale optimization problems.

\begin{figure}[h!]
    \centering
    \includegraphics[width=0.8\linewidth]{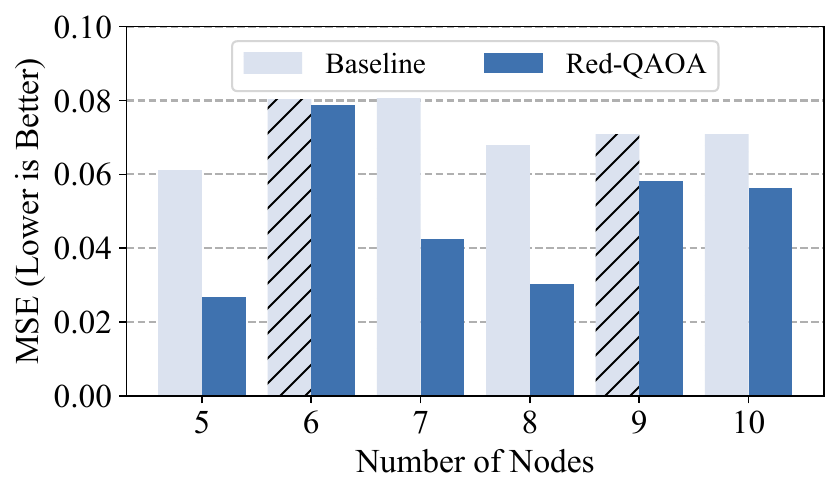}
    \caption{\ours outperforms noisy baseline QAOA on the 79-qubit Rigetti \textit{Aspen-M-3 device} by achieving a lower MSE across all evaluated cases.}
    \label{fig:rigetti_result}
\end{figure}

\noindent \textbf{Rigetti Device}: To further evaluate the performance of \ours on near-term quantum hardware, we employed it on the Rigetti Aspen-M-3 system with 79 qubits. Due to this device's higher error rates and limited access time, we benchmarked smaller graphs, 5 to 10 nodes, with a 1-layer QAOA. We compared the MSE between the ideal and noisy energy landscapes for \ours versus the baseline QAOA. As shown in Figure \ref{fig:rigetti_result}, \ours consistently achieved lower MSE across all evaluated cases on the Rigetti machine. By using a reduced QAOA circuit and obtaining enhanced performance, \ours showcases higher noise resilience even on today's noisy quantum devices.

\subsection{Varying Noise Models}
\begin{figure}[h]
    \centering
    \includegraphics[width=0.8\linewidth]{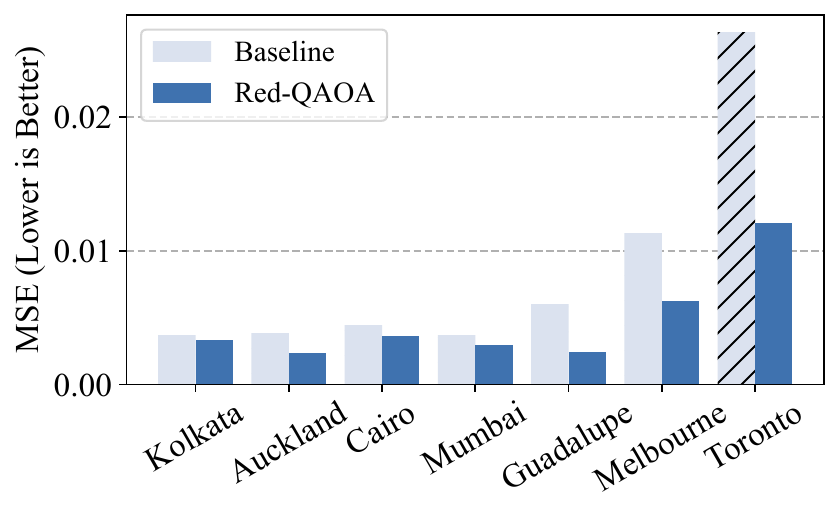}
    \caption{ours demonstrates enhanced noise tolerance versus baseline QAOA across devices with a wide range of error rates. MSE between the noise-free and noisy energy landscapes is consistently lower for \ours.}
    \label{fig:var_noise_model}
\end{figure}

To demonstrate the noise tolerance of \ours, we conducted an experiment using a random 10-node test graph and 1-layer QAOA with 1024 parameter sets. We calculated the mean squared error (MSE) between the noise-free energy landscape and landscapes generated under different noise models. The noise models were sampled from real IBM quantum device backends covering a wide range of error rates. As shown in Figure~\ref{fig:var_noise_model}, \ours consistently achieves a lower MSE than the baseline across all noise levels. This included noise models from the Kolkata backend, which has one of the lowest error rates among the existing IBM devices, and the retired Toronto device with substantially higher errors. By using a smaller QAOA circuit, \ours is inherently more tolerant of all types of noise.

\subsection{Increased Execution Throughput}

\begin{figure}[h!]
    \centering
    \includegraphics[width=0.9\linewidth]{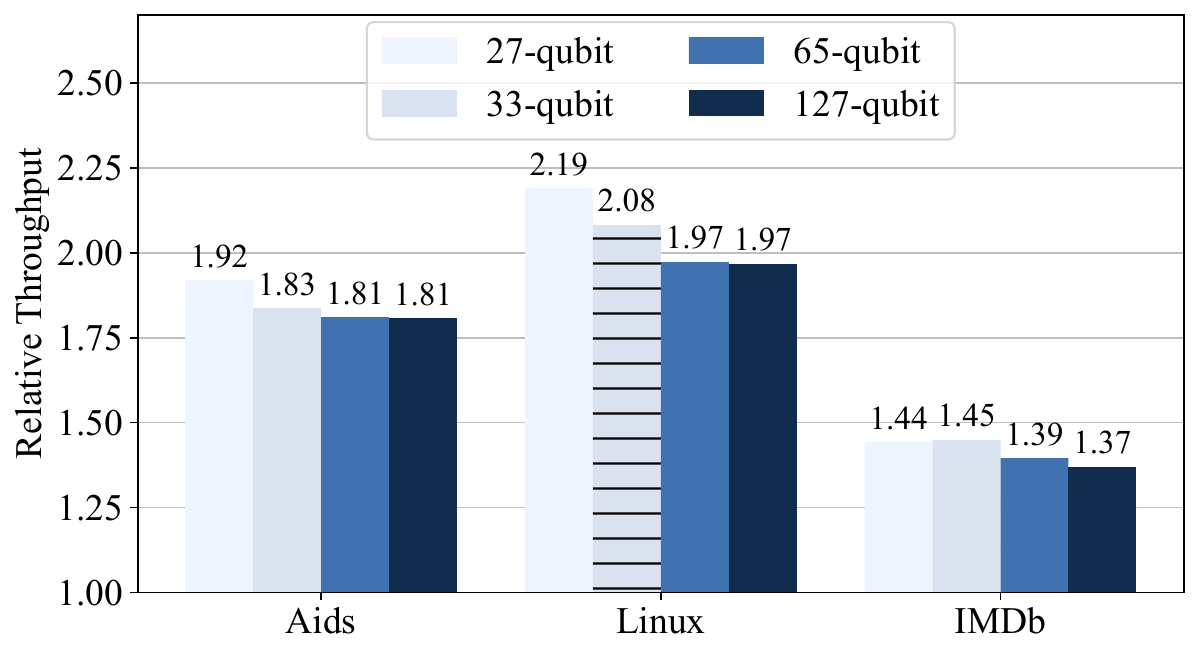}
    \caption{Expected throughput improvement of \ours compared to the baseline on 27, 33, 65, and 127-qubit devices using aids, Linux, and IMDb benchmarks. We observed significant improvements: around 1.85x for aids, 2.1x for Linux, and 1.4x for IMDb in terms of system throughput.}
    \label{fig:proj_throughput}
\end{figure}

Figure~\ref{fig:proj_throughput} shows the substantial throughput enhancement achieved with \ours, resulting in reduced execution time. We evaluated this improvement on large-scale quantum devices running multiple quantum circuits concurrently to optimize their utilization. The analyzed devices include Falcon 27-qubit, Eagle 33-qubit, Hummingbird 65-qubit, and Eagle 127-qubit. Notably, we observed throughput improvements of 1.85$\times$ for AIDS, 2.1$\times$ for Linux, and 1.4$\times$ for IMDb.
\section{Related Work}
\subsection{Parameter Transfer} 

Previous research~\cite{brandao_for_2018,galda_transferability_2021, shaydulin_parameter_2023} has observed similar findings regarding input instances. However, their typical approach involves using a set of small, computationally feasible input graphs and optimized parameters as a reference. When encountering a new input, they search the pre-optimized dataset, find the most similar graph, and apply its parameters directly. This approach for transferring optimal parameters between graphs has two significant limitations. Firstly, identifying transferability for a vast set of subgraphs becomes impractical due to the exponential growth of possible subgraph combinations with increasing graph size. Secondly, achieving mutual transfer conditions for large graphs is challenging, as it necessitates meeting the condition for every subgraph pair, limiting the concept's applicability. Therefore, these limitations raise doubts about the effectiveness and feasibility of existing theoretical frameworks for optimal QAOA parameter transferability.

Previous studies have shown transferability between regular graphs with uniform degree distribution. The precondition is automatically satisfied for two regular graphs with the same degree of parity. However, real-world data typically consists of irregular graphs. In particular, only \textbf{1.14\%} and \textbf{0\%} of the graphs from the AIDS and LINUX datasets that are used in this study are regular graphs respectively, while about \textbf{54\%} of the graphs in the IMDb dataset are regular.

\subsection{Enhancing QAOA Performance} 

Various warm start techniques are proposed to enhance quantum optimization algorithms by initializing them with educated guesses, leading to faster convergence and better solutions. Jain et al.~\cite{Jain_2022} proposed a graph neural network-based warm start for QAOA that is effective on various combinatorial optimization problems. CAFQA\cite{ravi2022cafqa} is a hybrid classical-quantum algorithm that improves solutions and convergence by finding parameters for a variational quantum algorithm. Egger et al.~\cite{Egger_2021} introduced warm-starting quantum optimization using classical relaxations of optimization problems and showed effectiveness in portfolio optimization and MAXCUT problems. Other works~\cite{li_2022, ayanzadeh2023frozenqubits, lao20212qan} have also utilized domain-specific knowledge of QAOA to enhance execution fidelity, while some works~\cite{zhuang2021efficient, Sun_2023} proposed efficient Hamiltonian transitions and expectation calculations.

Our work focuses on a complementary approach to these warm start and domain-specific techniques. By combining our approach with these methods, we can improve the performance of quantum optimization algorithms even further.

\subsection{Classical Optimization}

Using reduced or surrogate models to expedite optimization and learning is a well-established concept in classical domains. It has found applications in hyperparameter tuning for machine learning~\cite{doubly_stochastic}, approximating expensive black-box functions~\cite{high-dimensional}, discovering suitable initializations in topology optimization~\cite{Blind_Image}, serving as substitutes for costly real-world experiments~\cite{sontag2012tightening}, and facilitating policy transfer from simpler to more complex environments in reinforcement learning~\cite{Raiko2012DeepLM}. Additionally, the challenge posed by saddle points in high-dimensional non-convex optimization has led to the development of saddle-free Newton approaches to optimize neural networks more efficiently~\cite{dauphin2014identifying}. While these reduced model methods expedite optimization and transfer in complex domains, \ours is specifically tailored to address the challenges of quantum tasks.

\section{Conclusion}

In the NISQ era, noise-resilient techniques are crucial, especially for variational algorithms like Quantum Approximate Optimization Algorithm (QAOA). This paper introduces \ours, a framework designed to mitigate noise effects on QAOA in NISQ devices. \ours utilizes a reduced graph for parameter optimization, enhancing resilience to errors and yielding superior results. Experimental findings demonstrate a substantial improvement in the noise-affected energy landscape, resulting in improved outcomes.

This overall improvement can be attributed to \nodered reduction in node counts and \edgered reduction in edge counts. We tested \ours on real quantum devices and showcased promising results. Our findings highlight the efficacy of our methodology in optimizing quantum circuits, offering valuable insights for future research in quantum computing applications. By integrating our method with complementary techniques, we can further boost the efficiency and accuracy of quantum optimization algorithms, paving the way for new advancements in quantum computing.

\begin{acks}
This work was supported by the National Research Council (NRC) Canada grant AQC 003, AQC 213, and the Natural Sciences and Engineering Research Council of Canada (NSERC) [funding number RGPIN-2019-05059]. This material is also based upon work partially supported by the U.S. Department of Energy, Office of Science, National Quantum Information Science Research Centers, and Quantum Science Center. This research used resources from the Oak Ridge Leadership Computing Facility, a DOE Office of Science User Facility supported under Contract DE-AC05-00OR22725. This research also used resources of the National Energy Research Scientific Computing Center (NERSC), a U.S. Department of Energy Office of Science User Facility located at Lawrence Berkeley National Laboratory, operated under Contract No. DE-AC02-05CH11231. We thank Muqing Zheng for helping run device experiments on IBM Quantum devices.
\end{acks}

\appendix
\section{Artifact Appendix}

\subsection{Abstract}

Red-QAOA introduces an innovative approach to optimizing the Quantum Approximate Optimization Algorithm (QAOA) by substituting the standard QAOA circuit with a smaller, less error-prone version. This reduced circuit maintains similar optimal circuit parameters while offering enhanced efficiency and reduced computational errors, making it more suitable for execution on quantum devices and classical simulations. Red-QAOA is implemented in Python, utilizing the Qiskit framework for quantum circuit operations and NetworkX for graph-related computations. 

\subsection{Description}

\begin{itemize}
    \item \textbf{How to access:}\\
    \url{https://github.com/meng-ubc/Red-QAOA}
    \item \textbf{Hardware Dependencies:} The experiments can be accelerated using NVIDIA GPUs, though this is not a requirement for basic execution.
    \item \textbf{Software Dependencies:} The following Python packages are essential for running the experiments:
        \begin{itemize}
            \item \textbf{Qiskit:} Used for circuit simulation.
            \item \textbf{Networkx:} Provides graph-related utilities.
            \item \textbf{Scipy:} Employed for classical optimization tasks.
            \item \textbf{(Optional) torch-geometric:} This package offers Graph Neural Network (GNN) based graph pooling methods that are compared to Red-QAOA.
        \end{itemize}
    \item \textbf{Data Sets:} The following graph data sets are included within the repository and are utilized in our experiments:
        \begin{itemize}
            \item \textbf{Linux:} A dataset representing Linux kernel interaction networks.
            \item \textbf{AIDS:} A dataset involving molecular structures related to AIDS research.
            \item \textbf{IMDb:} Contains data from the IMDb movie database.
        \end{itemize}
    \item \textbf{Installation:} To install the necessary software for running the experiments, follow these steps:

        \begin{enumerate}
            \item \textbf{Python Version:} The scripts and tools have been tested with Python 3.11.
            \item \textbf{Required Packages:} A \texttt{requirements.txt} file is provided for installing required python packages.
             \item \textbf{Optional GPU Support:} For systems with CUDA capabilities, the optional package `qiskit-aer-gpu` can be installed separately to enable GPU acceleration. Note that the default `qiskit-aer` package has to be uninstalled before installing `qiskit-aer-gpu`.
        \end{enumerate}
         
\end{itemize}

\subsection{Experiment workflow}

\textbf{Running Experiments:} Each experiment is provided with a Python script. The script requires specific command-line arguments and can be executed as:

\begin{verbatim}
    python experiment_script.py [--arguments]
\end{verbatim} 

\noindent\textbf{Output and Analysis:} The script outputs the numerical result that can be directly compared to the numbers reported in the paper.

\subsection{Evaluation and expected results}

Our study employs two key metrics: Mean Square Error (MSE) and Approximation Ratio. 

\begin{itemize}
    \item \textbf{Mean Square Error (MSE)}: Utilized to measure the differences between two QAOA landscapes, this metric is applied in:
    \begin{enumerate}
        \item \textit{Ideal Execution with MSE}: Assessing the MSE between ideal QAOA landscapes.
        \item \textit{Noisy Execution with MSE}: Comparing the MSE in noisy QAOA landscapes to evaluate the impact of noise.
    \end{enumerate}

    \item \textbf{Approximation Ratio}: This metric assesses the end-to-end performance of Red-QAOA, with distinct objectives in ideal and noisy scenarios:
    \begin{enumerate}
        \item \textit{Ideal Execution with Approximation Ratio}: In ideal conditions, Red-QAOA aims to achieve results as close to the baseline QAOA performance.
    \end{enumerate}
\end{itemize}

In evaluating Red-QAOA, we have prepared a suite of experiments. Among these, three key experiments are crucial for understanding the efficacy and robustness of Red-QAOA, and they are:

\subsubsection{MSE Analysis of Red-QAOA under Noisy Execution}
This experiment reproduces Section 6.1 results in the paper.

\textbf{Script and Arguments:}
\begin{itemize}
    \item \textbf{Script:} \texttt{mse\_noisy.py}
    \item \textbf{Required Arguments:}
        \begin{itemize}
            \item \texttt{-n}: Specifies the number of nodes, ranging from 7 to 14, as used in the paper.
            \item \texttt{--width}: Sets the width of the landscape, defaulting to 32 (totalling 1024 executions).
            \item \texttt{--shots}: Defines the number of circuit executions, with a default of 8192.
            \item \texttt{--use\_gpu}: An optional flag to utilize GPU computing, with CPU as the default.
        \end{itemize}
\end{itemize}

\textbf{Result Analysis:}
The script produces two MSE values:
\begin{enumerate}
    \item The MSE between the noisy and ideal baseline landscapes.
    \item The MSE between the noisy Red-QAOA and ideal baseline landscapes.
\end{enumerate}

This analysis examines the relative difference between these MSE values. Due to the randomness in graph generation, absolute MSE values may vary; however, the focus should be on comparing the relative performance of Red-QAOA against the baselines under noisy conditions.

\subsubsection{MSE Analysis of Red-QAOA in Ideal Conditions}
This experiment reproduces results presented in Sections 6.2 and 6.3 in the paper.

\textbf{Script and Arguments:}
\begin{itemize}
    \item \textbf{Script:} \texttt{mse\_ideal.py}
    \item \textbf{Required Arguments:}
        \begin{itemize}
            \item \texttt{--graph\_set}: Specifies the graph dataset; options include aids, Linux, and IMDb.
            \item \texttt{--num\_graphs}: Defines the number of graphs for testing (it is recommended to test with at least ten graphs).
            \item \texttt{--p}: Sets the number of QAOA layers.
        \end{itemize}
    \item \textbf{Optional Arguments:}
        \begin{itemize}
            \item \texttt{--num\_points}: The number of points sampled for the landscape, defaulting to 1024.
            \item \texttt{--shots}: The number of shots for circuit execution, with a default of 8192.
            \item \texttt{--use\_gpu}: Indicates whether to use a GPU backend; the default is CPU.
            \item \texttt{--min\_nodes} and \texttt{--max\_nodes}: Specifies the range of nodes, defaulting to 0 to 10 for Section 6.2 and 10 to 20 for Section 6.3.
        \end{itemize}
\end{itemize}

\textbf{Interpreting Results:}
The output should include MSE values along with node and edge reductions, which can be directly compared with the figures and data presented in the paper. 

\subsubsection{End-to-End Performance Evaluation of Red-QAOA in Ideal Conditions}

This experiment evaluates the end-to-end performance of the Red-QAOA under ideal, noise-free conditions, which is discussed in Section 6.4.1 in the paper.

\textbf{Script and Arguments:}
\begin{itemize}
    \item \textbf{Script:} \texttt{end\_to\_end.py}
    
    \item \textbf{Required Argument:}
        \begin{itemize}
            \item \texttt{--p}: Specifies the number of QAOA layers.
        \end{itemize}

    \item \textbf{Optional Arguments:}
        \begin{itemize}
            \item \texttt{--num\_graphs}: Sets the number of graphs for testing, with a default of 100.
            \item \texttt{--num\_nodes}: Determines the number of nodes in each graph; the default is 30 to align with the paper's setting. A smaller value, like 10, is suggested for reduced computational overhead.
            \item \texttt{--shots}: Number of shots for circuit execution, defaulting to 8192.
            \item \texttt{--use\_gpu}: Allows the option to use a GPU backend, with CPU as the default.
        \end{itemize}
\end{itemize}

\textbf{Interpreting Results:}
The script conducts multiple optimization restarts for each test case. It reports the average optimization result for Red-QAOA across all restarts, comparing these to the baseline and the optimal results. These values can be directly compared with those reported in the paper to validate Red-QAOA's performance in ideal scenarios.

\subsubsection{Reproducing the Figures}
The figures in the paper are generated using matplotlib package. The repository contains documentation (README\_plot.md) to reproduce the corresponding figures in the paper.

\subsection{Experiment Customization}
The experiments in our paper, including variations like different numbers of QAOA layers, are designed with flexibility in mind. Key experiment parameters are set as required arguments in our scripts, ensuring consistency with the study's main findings. Additionally, a range of optional arguments is available, allowing for fine-tuning and more detailed, sensitive testing scenarios.

\bibliographystyle{unsrt}
\balance
\bibliography{references}

\begin{thebibliography}{10}

\bibitem{supremacy}
Frank Arute, Kunal Arya, Ryan Babbush, Dave Bacon, Joseph Bardin, Rami Barends, et~al.
\newblock Quantum supremacy using a programmable superconducting processor, 2019.

\bibitem{Bruzewicz2019}
Colin~D. Bruzewicz, John Chiaverini, Robert McConnell, and Jeremy~M. Sage.
\newblock Trapped-ion quantum computing: Progress and challenges.
\newblock {\em Applied Physics Reviews}, 6, 2019.

\bibitem{Cao2019}
Yudong Cao, Jonathan Romero, Jonathan~P Olson, Matthias Degroote, Peter~D Johnson, M{\'a}ria Kieferov{\'a}, Ian~D Kivlichan, Tim Menke, Borja Peropadre, Nicolas~PD Sawaya, et~al.
\newblock Quantum chemistry in the age of quantum computing.
\newblock {\em Chemical reviews}, 119(19):10856--10915, 2019.

\bibitem{Farhi2001}
E.~Farhi, J.~Goldstone, S.~Gutmann, J.~Lapan, A.~Lundgren, and D.~Preda.
\newblock A quantum adiabatic evolution algorithm applied to random instances of an np-complete problem.
\newblock {\em Science}, 292, 2001.

\bibitem{Feynman1982}
Richard~P. Feynman.
\newblock Simulating physics with computers.
\newblock {\em International Journal of Theoretical Physics}, 21, 1982.

\bibitem{ibm_roadmap}
Jay Gambetta.
\newblock {Expanding the IBM Quantum roadmap to anticipate the future of quantum-centric supercomputing}.
\newblock \url{https://research.ibm.com/blog/ibm-quantum-roadmap-2025}, 2022.
\newblock [Online; accessed 1-April-2023].

\bibitem{grover_fast_1996}
Lov~K Grover.
\newblock A fast quantum mechanical algorithm for database search.
\newblock {\em arXiv preprint quant-ph/9605043}, 1996.

\bibitem{shor_polynomial-time_1999}
Peter~W Shor.
\newblock Polynomial-time algorithms for prime factorization and discrete logarithms on a quantum computer.
\newblock {\em SIAM review}, 41(2):303--332, 1999.
\newblock Publisher: SIAM.

\bibitem{dasmultiprogramming}
Poulami Das, Swamit~S. Tannu, Prashant~J. Nair, and Moinuddin Qureshi.
\newblock A case for multi-programming quantum computers.
\newblock MICRO '52, page 291–303, New York, NY, USA, 2019. Association for Computing Machinery.

\bibitem{farhi_quantum_2014}
Edward Farhi, Jeffrey Goldstone, and Sam Gutmann.
\newblock A quantum approximate optimization algorithm.
\newblock {\em arXiv preprint arXiv:1411.4028}, 2014.

\bibitem{fletcher_practical_2000}
R.~Fletcher.
\newblock {\em Practical {Methods} of {Optimization}: {Fletcher}/{Practical} {Methods} of {Optimization}}.
\newblock John Wiley \& Sons, Ltd, Chichester, West Sussex England, May 2000.

\bibitem{lotshaw_empirical_2021}
Phillip~C. Lotshaw, Travis~S. Humble, Rebekah Herrman, James Ostrowski, and George Siopsis.
\newblock Empirical performance bounds for quantum approximate optimization.
\newblock {\em Quantum Information Processing}, 20(12):403, December 2021.

\bibitem{wang_quantum_2018}
Zhihui Wang, Stuart Hadfield, Zhang Jiang, and Eleanor~G. Rieffel.
\newblock Quantum approximate optimization algorithm for {MaxCut}: {A} fermionic view.
\newblock {\em Physical Review A}, 97(2):022304, February 2018.

\bibitem{ward_qaoa_2018}
Jonathan Ward, Johannes Otterbach, Gavin Crooks, Nicholas Rubin, and Marcus da~Silva.
\newblock {QAOA} {Performance} {Benchmarks} using {Max}-{Cut}.
\newblock In {\em {APS} {Meeting} {Abstracts}}, 2018.

\bibitem{zhou_quantum_2020}
Leo Zhou, Sheng-Tao Wang, Soonwon Choi, Hannes Pichler, and Mikhail~D. Lukin.
\newblock Quantum {Approximate} {Optimization} {Algorithm}: {Performance}, {Mechanism}, and {Implementation} on {Near}-{Term} {Devices}.
\newblock {\em Physical Review X}, 10(2):021067, June 2020.

\bibitem{cerezo_variational_2021}
M.~Cerezo, Andrew Arrasmith, Ryan Babbush, Simon~C. Benjamin, Suguru Endo, Keisuke Fujii, Jarrod~R. McClean, Kosuke Mitarai, Xiao Yuan, Lukasz Cincio, and Patrick~J. Coles.
\newblock Variational {Quantum} {Algorithms}.
\newblock {\em Nature Reviews Physics}, 3(9):625--644, August 2021.
\newblock arXiv:2012.09265 [quant-ph, stat].

\bibitem{endo_variational_2020}
Suguru Endo, Jinzhao Sun, Ying Li, Simon~C. Benjamin, and Xiao Yuan.
\newblock Variational {Quantum} {Simulation} of {General} {Processes}.
\newblock {\em Physical Review Letters}, 125(1):010501, June 2020.

\bibitem{huang_near-term_2022}
He-Liang Huang, Xiao-Yue Xu, Chu Guo, Guojing Tian, Shi-Jie Wei, Xiaoming Sun, Wan-Su Bao, and Gui-Lu Long.
\newblock Near-{Term} {Quantum} {Computing} {Techniques}: {Variational} {Quantum} {Algorithms}, {Error} {Mitigation}, {Circuit} {Compilation}, {Benchmarking} and {Classical} {Simulation}, December 2022.
\newblock arXiv:2211.08737 [quant-ph].

\bibitem{mengquantumsim}
Meng Wang, Fei Hua, Chenxu Liu, Nicholas Bauman, Karol Kowalski, Daniel Claudino, Travis Humble, Prashant Nair, and Ang Li.
\newblock Enabling scalable vqe simulation on leading hpc systems.
\newblock In {\em Proceedings of the SC '23 Workshops of The International Conference on High Performance Computing, Network, Storage, and Analysis}, SC-W '23, page 1460–1467, New York, NY, USA, 2023. Association for Computing Machinery.

\bibitem{wang2022tqsim}
Meng Wang, Rui Huang, Swamit Tannu, and Prashant Nair.
\newblock Tqsim: a case for reuse-focused tree-based quantum circuit simulation.
\newblock {\em arXiv preprint arXiv:2203.13892}, 2022.

\bibitem{mengrestart}
Meng Wang, Bo~Fang, Ang Li, and Prashant Nair.
\newblock Efficient qaoa optimization using directed restarts and graph lookup.
\newblock In {\em Proceedings of the 2023 International Workshop on Quantum Classical Cooperative}, QCCC '23, page 5–8, New York, NY, USA, 2023. Association for Computing Machinery.

\bibitem{tannuecc}
Swamit~S. Tannu, Zachary~A. Myers, Prashant~J. Nair, Douglas~M. Carmean, and Moinuddin~K. Qureshi.
\newblock Taming the instruction bandwidth of quantum computers via hardware-managed error correction.
\newblock In {\em Proceedings of the 50th Annual IEEE/ACM International Symposium on Microarchitecture}, MICRO-50 '17, page 679–691, New York, NY, USA, 2017. Association for Computing Machinery.

\bibitem{tannunisq}
Swamit~S. Tannu and Moinuddin~K. Qureshi.
\newblock Not all qubits are created equal: A case for variability-aware policies for nisq-era quantum computers.
\newblock In {\em Proceedings of the Twenty-Fourth International Conference on Architectural Support for Programming Languages and Operating Systems}, ASPLOS '19, page 987–999, New York, NY, USA, 2019. Association for Computing Machinery.

\bibitem{tannumeasurement}
Swamit~S. Tannu and Moinuddin~K. Qureshi.
\newblock Mitigating measurement errors in quantum computers by exploiting state-dependent bias.
\newblock In {\em Proceedings of the 52nd Annual IEEE/ACM International Symposium on Microarchitecture}, MICRO '52, page 279–290, New York, NY, USA, 2019. Association for Computing Machinery.

\bibitem{tannudissimilar}
Swamit~S. Tannu and Moinuddin Qureshi.
\newblock Ensemble of diverse mappings: Improving reliability of quantum computers by orchestrating dissimilar mistakes.
\newblock In {\em Proceedings of the 52nd Annual IEEE/ACM International Symposium on Microarchitecture}, MICRO '52, page 253–265, New York, NY, USA, 2019. Association for Computing Machinery.

\bibitem{brandao_for_2018}
Fernando G. S.~L. Brandao, Michael Broughton, Edward Farhi, Sam Gutmann, and Hartmut Neven.
\newblock For {Fixed} {Control} {Parameters} the {Quantum} {Approximate} {Optimization} {Algorithm}'s {Objective} {Function} {Value} {Concentrates} for {Typical} {Instances}, December 2018.
\newblock arXiv:1812.04170 [quant-ph].

\bibitem{shaydulin_qaoakit_2021}
Ruslan Shaydulin, Kunal Marwaha, Jonathan Wurtz, and Phillip~C. Lotshaw.
\newblock {QAOAKit}: {A} {Toolkit} for {Reproducible} {Study}, {Application}, and {Verification} of the {QAOA}.
\newblock In {\em 2021 {IEEE}/{ACM} {Second} {International} {Workshop} on {Quantum} {Computing} {Software} ({QCS})}, pages 64--71, November 2021.
\newblock arXiv:2110.05555 [quant-ph].

\bibitem{galda_transferability_2021}
Alexey Galda, Xiaoyuan Liu, Danylo Lykov, Yuri Alexeev, and Ilya Safro.
\newblock Transferability of optimal {QAOA} parameters between random graphs, June 2021.
\newblock arXiv:2106.07531 [quant-ph].

\bibitem{shaydulin_parameter_2023}
Ruslan Shaydulin, Phillip~C. Lotshaw, Jeffrey Larson, James Ostrowski, and Travis~S. Humble.
\newblock Parameter {Transfer} for {Quantum} {Approximate} {Optimization} of {Weighted} {MaxCut}.
\newblock {\em ACM Transactions on Quantum Computing}, page 3584706, February 2023.
\newblock arXiv:2201.11785 [quant-ph].

\bibitem{dasimitation}
Poulami Das, Eric Kessler, and Yunong Shi.
\newblock The imitation game: Leveraging copycats for robust native gate selection in nisq programs.
\newblock In {\em 2023 IEEE International Symposium on High-Performance Computer Architecture (HPCA)}, pages 787--801, 2023.

\bibitem{dasadapt}
Poulami Das, Swamit Tannu, Siddharth Dangwal, and Moinuddin Qureshi.
\newblock Adapt: Mitigating idling errors in qubits via adaptive dynamical decoupling.
\newblock In {\em MICRO-54: 54th Annual IEEE/ACM International Symposium on Microarchitecture}, MICRO '21, page 950–962, New York, NY, USA, 2021. Association for Computing Machinery.

\bibitem{dangwalvarsaw}
Siddharth Dangwal, Gokul~Subramanian Ravi, Poulami Das, Kaitlin~N. Smith, Jonathan~Mark Baker, and Frederic~T. Chong.
\newblock Varsaw: Application-tailored measurement error mitigation for variational quantum algorithms.
\newblock In {\em Proceedings of the 28th ACM International Conference on Architectural Support for Programming Languages and Operating Systems, Volume 4}, ASPLOS '23, page 362–377, New York, NY, USA, 2024. Association for Computing Machinery.

\bibitem{tannuhammer}
Swamit Tannu, Poulami Das, Ramin Ayanzadeh, and Moinuddin Qureshi.
\newblock Hammer: boosting fidelity of noisy quantum circuits by exploiting hamming behavior of erroneous outcomes.
\newblock In {\em Proceedings of the 27th ACM International Conference on Architectural Support for Programming Languages and Operating Systems}, ASPLOS '22, page 529–540, New York, NY, USA, 2022. Association for Computing Machinery.

\bibitem{deo_graph_2016}
Narsingh Deo.
\newblock {\em Graph theory with applications to engineering and computer science}.
\newblock Dover Publications, Inc, Mineola New York, 2016.

\bibitem{ranjan_asap_2020}
Ekagra Ranjan, Soumya Sanyal, and Partha~Pratim Talukdar.
\newblock {ASAP}: {Adaptive} {Structure} {Aware} {Pooling} for {Learning} {Hierarchical} {Graph} {Representations}, February 2020.
\newblock arXiv:1911.07979 [cs, stat].

\bibitem{knyazev_understanding_2019}
Boris Knyazev, Graham~W. Taylor, and Mohamed~R. Amer.
\newblock Understanding {Attention} and {Generalization} in {Graph} {Neural} {Networks}, October 2019.
\newblock arXiv:1905.02850 [cs, stat].

\bibitem{lee_self-attention_2019}
Junhyun Lee, Inyeop Lee, and Jaewoo Kang.
\newblock Self-{Attention} {Graph} {Pooling}, June 2019.
\newblock arXiv:1904.08082 [cs, stat].

\bibitem{cangea_towards_2018}
Cătălina Cangea, Petar Veličković, Nikola Jovanović, Thomas Kipf, and Pietro Liò.
\newblock Towards {Sparse} {Hierarchical} {Graph} {Classifiers}, November 2018.
\newblock arXiv:1811.01287 [cs, stat].

\bibitem{gao_graph_2019}
Hongyang Gao and Shuiwang Ji.
\newblock Graph {U}-{Nets}, May 2019.
\newblock arXiv:1905.05178 [cs, stat].

\bibitem{Lotshaw_2021}
Phillip~C. Lotshaw, Travis~S. Humble, Rebekah Herrman, James Ostrowski, and George Siopsis.
\newblock Empirical performance bounds for quantum approximate optimization.
\newblock {\em Quantum Information Processing}, 20(12), nov 2021.

\bibitem{kirkpatrick_optimization_1983}
S.~Kirkpatrick, C.~D. Gelatt, and M.~P. Vecchi.
\newblock Optimization by {Simulated} {Annealing}.
\newblock {\em Science}, 220(4598):671--680, May 1983.

\bibitem{laarhoven_simulated_1992}
Peter J. M.~van Laarhoven, Peter J. M.~van Laarhoven, and Emile H.~L. Aarts.
\newblock {\em Simulated annealing: theory and applications}.
\newblock Number~37 in Mathematics and its applications {\textless}{Dordrecht}{\textgreater}. Kluwer, Dordrecht, reprinted with corr. 1988, reprinted edition, 1992.

\bibitem{da_vitoria_lobo_iam_2008}
Kaspar Riesen and Horst Bunke.
\newblock {IAM} {Graph} {Database} {Repository} for {Graph} {Based} {Pattern} {Recognition} and {Machine} {Learning}.
\newblock In Niels da~Vitoria~Lobo, Takis Kasparis, Fabio Roli, James~T. Kwok, Michael Georgiopoulos, Georgios~C. Anagnostopoulos, and Marco Loog, editors, {\em Structural, {Syntactic}, and {Statistical} {Pattern} {Recognition}}, volume 5342, pages 287--297. Springer Berlin Heidelberg, Berlin, Heidelberg, 2008.
\newblock Series Title: Lecture Notes in Computer Science.

\bibitem{wang_efficient_2012}
Xiaoli Wang, Xiaofeng Ding, Anthony~K.H. Tung, Shanshan Ying, and Hai Jin.
\newblock An {Efficient} {Graph} {Indexing} {Method}.
\newblock In {\em 2012 {IEEE} 28th {International} {Conference} on {Data} {Engineering}}, pages 210--221, Arlington, VA, USA, April 2012. IEEE.

\bibitem{yanardag_deep_2015}
Pinar Yanardag and S.V.N. Vishwanathan.
\newblock Deep {Graph} {Kernels}.
\newblock In {\em Proceedings of the 21th {ACM} {SIGKDD} {International} {Conference} on {Knowledge} {Discovery} and {Data} {Mining}}, pages 1365--1374, Sydney NSW Australia, August 2015. ACM.

\bibitem{erdHos1960evolution}
Paul Erd{\H{o}}s, Alfr{\'e}d R{\'e}nyi, et~al.
\newblock On the evolution of random graphs.
\newblock {\em Publ. Math. Inst. Hung. Acad. Sci}, 5(1):17--60, 1960.

\bibitem{Qiskit}
{Qiskit contributors}.
\newblock Qiskit: An open-source framework for quantum computing, 2023.

\bibitem{dasjigsaw}
Poulami Das, Swamit Tannu, and Moinuddin Qureshi.
\newblock Jigsaw: Boosting fidelity of nisq programs via measurement subsetting.
\newblock In {\em MICRO-54: 54th Annual IEEE/ACM International Symposium on Microarchitecture}, MICRO '21, page 937–949, New York, NY, USA, 2021. Association for Computing Machinery.

\bibitem{li2019tackling}
Gushu Li, Yufei Ding, and Yuan Xie.
\newblock Tackling the qubit mapping problem for nisq-era quantum devices, 2019.

\bibitem{Fey/Lenssen/2019}
Matthias Fey and Jan~E. Lenssen.
\newblock Fast graph representation learning with {PyTorch Geometric}.
\newblock In {\em ICLR Workshop on Representation Learning on Graphs and Manifolds}, 2019.

\bibitem{wack2021quality}
Andrew Wack, Hanhee Paik, Ali Javadi-Abhari, Petar Jurcevic, Ismael Faro, Jay~M. Gambetta, and Blake~R. Johnson.
\newblock Quality, speed, and scale: three key attributes to measure the performance of near-term quantum computers, 2021.

\bibitem{powell1994direct}
Michael~JD Powell.
\newblock A direct search optimization method that models the objective and constraint functions by linear interpolation.
\newblock In {\em Advances in optimization and numerical analysis}, pages 51--67. Springer, 1994.

\bibitem{alam2020circuit}
Mahabubul Alam, Abdullah Ash-Saki, and Swaroop Ghosh.
\newblock Circuit compilation methodologies for quantum approximate optimization algorithm.
\newblock In {\em 2020 53rd Annual IEEE/ACM International Symposium on Microarchitecture (MICRO)}, pages 215--228. IEEE, 2020.

\bibitem{ayanzadeh2023frozenqubits}
Ramin Ayanzadeh, Narges Alavisamani, Poulami Das, and Moinuddin Qureshi.
\newblock Frozenqubits: Boosting fidelity of qaoa by skipping hotspot nodes, 2023.

\bibitem{temme2017error}
Kristan Temme, Sergey Bravyi, and Jay~M Gambetta.
\newblock Error mitigation for short-depth quantum circuits.
\newblock {\em Physical review letters}, 119(18):180509, 2017.

\bibitem{Jain_2022}
Nishant Jain, Brian Coyle, Elham Kashefi, and Niraj Kumar.
\newblock Graph neural network initialisation of quantum approximate optimisation.
\newblock {\em Quantum}, 6:861, nov 2022.

\bibitem{ravi2022cafqa}
Gokul~Subramanian Ravi, Pranav Gokhale, Yi~Ding, William~M. Kirby, Kaitlin~N. Smith, Jonathan~M. Baker, Peter~J. Love, Henry Hoffmann, Kenneth~R. Brown, and Frederic~T. Chong.
\newblock Cafqa: A classical simulation bootstrap for variational quantum algorithms, 2022.

\bibitem{Egger_2021}
Daniel~J. Egger, Jakub Mare{\v{c} }ek, and Stefan Woerner.
\newblock Warm-starting quantum optimization.
\newblock {\em Quantum}, 5:479, jun 2021.

\bibitem{li_2022}
Gushu Li, Anbang Wu, Yunong Shi, Ali Javadi-Abhari, Yufei Ding, and Yuan Xie.
\newblock Paulihedral: A generalized block-wise compiler optimization framework for quantum simulation kernels.
\newblock In {\em Proceedings of the 27th ACM International Conference on Architectural Support for Programming Languages and Operating Systems}, ASPLOS '22, page 554–569, New York, NY, USA, 2022. Association for Computing Machinery.

\bibitem{lao20212qan}
Lingling Lao and Dan~E. Browne.
\newblock 2qan: A quantum compiler for 2-local qubit hamiltonian simulation algorithms, 2021.

\bibitem{zhuang2021efficient}
Wei-Feng Zhuang, Ya-Nan Pu, Hong-Ze Xu, Xudan Chai, Yanwu Gu, Yunheng Ma, Shahid Qamar, Chen Qian, Peng Qian, Xiao Xiao, Meng-Jun Hu, and Dong~E. Liu.
\newblock Efficient classical computation of quantum mean values for shallow qaoa circuits, 2021.

\bibitem{Sun_2023}
Zheng-Hang Sun, Yong-Yi Wang, Jian Cui, and Heng Fan.
\newblock Improving the performance of quantum approximate optimization for preparing non-trivial quantum states without translational symmetry.
\newblock {\em New Journal of Physics}, 25(1):013015, jan 2023.

\bibitem{doubly_stochastic}
Ron Zass and Amnon Shashua.
\newblock Doubly stochastic normalization for spectral clustering.
\newblock In B.~Sch\"{o}lkopf, J.~Platt, and T.~Hoffman, editors, {\em Advances in Neural Information Processing Systems}, volume~19. MIT Press, 2006.

\bibitem{high-dimensional}
Pradeep Ravikumar, Martin~J. Wainwright, and John~D. Lafferty.
\newblock {High-dimensional Ising model selection using l1-regularized logistic regression}.
\newblock {\em The Annals of Statistics}, 38(3):1287 -- 1319, 2010.

\bibitem{Blind_Image}
Tal Kenig, Zvi Kam, and Arie Feuer.
\newblock Blind image deconvolution using machine learning for three-dimensional microscopy.
\newblock {\em IEEE Transactions on Pattern Analysis and Machine Intelligence}, 32(12):2191--2204, 2010.

\bibitem{sontag2012tightening}
David Sontag, Talya Meltzer, Amir Globerson, Tommi~S. Jaakkola, and Yair Weiss.
\newblock Tightening lp relaxations for map using message passing, 2012.

\bibitem{Raiko2012DeepLM}
Tapani Raiko, Harri Valpola, and Yann LeCun.
\newblock Deep learning made easier by linear transformations in perceptrons.
\newblock In {\em International Conference on Artificial Intelligence and Statistics}, 2012.

\bibitem{dauphin2014identifying}
Yann Dauphin, Razvan Pascanu, Caglar Gulcehre, Kyunghyun Cho, Surya Ganguli, and Yoshua Bengio.
\newblock Identifying and attacking the saddle point problem in high-dimensional non-convex optimization, 2014.

\end{thebibliography}

\end{document}